\documentclass[12pt]{article}

\newcommand{\bbr}{I\!\! R}
\newcommand{\bbz}{Z\!\!\! Z}

\newcommand{\2}{$^2$}
\newcommand{\3}{$^3$}
\newcommand{\4}{$_4$}
\newcommand{\5}{$_5$}
\newcommand{\x}{arXiv:}

\usepackage{graphicx}

\begin{document}
\thispagestyle{empty}
\begin{center}

\null \vskip-1truecm \vskip2truecm {\bf Orbifold Physics and de
Sitter Spacetime\\} \vskip2truecm Brett McInnes \vskip2truecm

National University of Singapore\\
email: matmcinn@nus.edu.sg\\

\end{center}
\vskip1truecm \centerline{ABSTRACT} \baselineskip=15pt
\medskip

It is
generally believed the way to resolve the black hole information
paradox in string theory is to embed the black hole in
anti-deSitter spacetime --- \emph{without} of course claiming that
Schwarzschild-AdS is a realistic spacetime. Here we propose that,
similarly, the best way to study topologically non-trivial versions of de Sitter spacetime from a stringy
point of view is to
embed them in an anti-de Sitter orbifold bulk, again \emph{without}
claiming that this is literally how de Sitter arises in string
theory. Our results indicate that string theory may rule out the
more complex spacetime topologies which are compatible with local
de Sitter geometry, while still allowing the simplest versions.

\vskip3.5truecm
\begin{center}

\end{center}

\newpage

\addtocounter{section}{1}
\section*{1. AdS Embedding as a Toolbox }

It is well known that it is difficult to obtain realistic curved
spacetimes from string theory. This is not necessarily a drawback
in itself, but it does impede the analysis of important questions
which string theory might be expected to answer. Ingenious ways of
answering these questions have been found, without actually
exhibiting the desired spacetimes explicitly.

An outstanding example of this is the black hole information
paradox. We do not yet know how to follow the detailed evolution
of a realistic evaporating black hole in string theory (but see
\cite{kn:mathur} and its references for some ideas), but it is
generally believed that string theory predicts that there is no
information loss. The strongest argument in this direction is due
to Maldacena \cite{kn:maldacena}: one simply embeds the black hole
in AdS\4, using the Schwarzschild-AdS\4 metric, and shows that a
dual field theory can be constructed in the familiar AdS/CFT
style. Since the evolution of the field theory is unitary, it is
clear that the same should be true of the black hole evaporation
process. Notice that \emph{no claim is made that
Schwarzschild}-AdS\4 \emph{is a realistic spacetime}; in fact of
course a real black hole spacetime is utterly different.
Nevertheless one believes that AdS is so well-understood in string
theory that one can reasonably expect to use it to learn something
about \emph{generic} string black hole evaporation in this manner.

In a similar way, we do not yet know exactly how accelerating
spacetimes can arise within string theory (see \cite{kn:balasub}
for a review of the current situation). But it is now clear
that our Universe is accelerating \cite{kn:carroll}, and so string theory should help us to answer some of the many
questions that arise regarding the quantum nature of an
accelerating cosmos. By analogy with the black hole information
paradox, we might hope that some clues --- if not complete answers
--- can be found by \emph{formally} embedding a candidate accelerating cosmology in a
suitable version of AdS\5. As in the black hole case, we shall
\emph{not} claim that accelerating spacetimes can be literally
obtained as a brane-world in string theory; as in the black hole
case, we accept that a real derivation of de Sitter spacetime from
string theory will certainly be far more complex than that. We
regard the AdS embedding as a \emph{tool} which may allow us to
make progress pending a truly ``stringy" derivation of cosmic
acceleration. (For the more literal interpretation, not
necessarily in the string context, see
\cite{kn:zerbini}\cite{kn:branenewworld}\cite{kn:maartens}.)

In fact, dS\4 embeds in AdS\5 in an extremely natural way: we do
not even have to modify the local geometry of AdS\5, as we do in
the black hole case. This is particularly clear in the Euclidean
formulation, since the five-dimensional hyperbolic space H$^5$
(Euclidean AdS\5) metric with curvature $-$1/L$^2$ is
\begin{equation} \label{eq:one}
g(H^5) =  dr \otimes dr + sinh^2(r/L)g(S^4),
\end{equation}
where $g$(S$^4$) is the metric of the 4-sphere (``Euclidean
deSitter") of curvature +1/L$^2$. (In an attempt to avoid
confusion, we shall throughout this work indicate the dimensions
of Euclidean spaces by superscripts, and those of Lorentzian
spaces by subscripts.) When we embed dS\4 or related spacetimes in versions of AdS\5 in this way,
we shall refer to a de Sitter \emph{slice} of anti-de Sitter, avoiding the term \emph{brane} to emphasise
that we are not committed to a literal brane-world scenario.

What kinds of questions about accelerating universes might we hope
to answer by means of a formal AdS\5 embedding? One example, which
has in fact already been analysed \cite{kn:HMS} in exactly this
way, is the question of the \emph{entropy of de Sitter spacetime}.
Related examples might be the class of problems (see for example
\cite{kn:susskind}) which arise when one tries to extend the
holographic principle to de Sitter spacetime
\cite{kn:witten1}\cite{kn:strominger}\cite{kn:gao1}\cite{kn:me}\cite{kn:larsen}\cite{kn:od1}\cite{kn:od2}\cite{kn:staruszkiewicz}
\cite{kn:mcnees}\cite{kn:schaar}\cite{kn:eva}\cite{kn:setare}.
Since holography works so well \cite{kn:magoo} for AdS\5, it is
natural to try to investigate de Sitter holography by embedding
dS\4 in AdS\5, and one of our objectives here is to set the stage
for this. In the Lorentzian case, the conformal boundary of the
embedded dS\4 {\em actually lies on the conformal boundary of the
ambient AdS$_5$}. In view of the AdS/CFT correspondence, this is
clearly an important point. (It appears in \cite{kn:branenewworld}
--- see figure 2 of that reference  --- but is not further
analysed.)

Another question which can be addressed by means of a formal AdS\5
embedding is that of the \emph{topology} of de Sitter spacetime.
It is not generally appreciated that there are in fact many
different topological spaces which can accept the de Sitter metric
\emph{locally}. We have no reason to believe that the version of
de Sitter spacetime which may emerge, possibly in truncated or
metastable form (see \cite{kn:trivedi}) from string theory, will
necessarily be the most familiar version with symmetry group
O(1,4). In fact, just the opposite is true, since a truncated or
otherwise mutilated spacetime will not be maximally symmetric. Furthermore,
Witten \cite{kn:witten1} has stressed that \emph{there is every reason to
doubt that quantum gravity effects related to the acceleration of
the universe are symmetrical under the full de Sitter group}
O(1,4). This doubt will be confirmed if, as has been suggested \cite{kn:banks1}, the entropy of de Sitter spacetime arises
ultimately from a finite-dimensional Hilbert space of states,
since the full de Sitter group has no finite-dimensional unitary
representations. This suggests that we should study the \emph{less
symmetrical} versions of de Sitter spacetime, many of which have
compact symmetry groups. There are various possible kinds of
``less-symmetrical" de Sitter spacetime; the version
\cite{kn:louko}\cite{kn:mcinnes} which differs least from
``ordinary" de Sitter spacetime is obtained by taking the spatial
sections to be copies of the real projective space $\bbr$P$^3$
instead of the three-sphere S$^3$, while the time axis is left
untouched. We call this dS($\bbr$P$^3$); it is a four-dimensional
spacetime with only six Killing vectors. (We call the spherical
version dS(S$^3$), and use dS$_4$ when we do not wish to be
specific.) A much more radical departure from conventional
cosmology is to pass to ``elliptic" de Sitter (see
\cite{kn:parikh}\cite{kn:aguirre}); in this case, the
identification involves both space and time, which entails a loss
of time-orientability. There are still other possibilities, which
we shall introduce in this work. The question then
is: what are the physical consequences of modifying the topology
of de Sitter spacetime? As we shall see, this is a question on
which a formal AdS\5 embedding can shed much light.

In accordance with our general ``AdS toolbox" philosophy, we
advocate that these variants of dS\4 should be studied by
embedding them in a suitable version of AdS\5. We shall argue that
the modifications of AdS\5 which are necessary to achieve this
often (though not always) lead to serious instabilities. In this
way, the AdS\5 embedding strongly suggests that string theory
severely constrains the topology of an accelerating universe.

We start with a discussion of the ambiguities in the global
structures of de Sitter and anti-de Sitter spacetimes, beginning
with the maximally symmetric versions. This is followed by an
explanation of how it is possible for the boundary of dS$_4$ to
``touch"  that of AdS$_5$. Then we
shall explain the consequences of taking the quotients of both
AdS$_5$ and dS$_4$ by various small, finite groups of isometries,
so that the symmetry groups are reduced. Next we discuss the
embedding of various versions of ``de Sitter spacetime" in
suitable versions of ``anti-de Sitter spacetime". Finally, recent
results on tachyonic instabilities in AdS orbifolds, combined with
a study of the breaking of supersymmetry, allow us to constrain
the global geometries of both. Inter alia we gain some insights
into de Sitter holography.

\addtocounter{section}{1}
\section*{2. Maximally Symmetric Versions of [Anti]de Sitter}

As is well known, there are many spaces with the local geometry of
anti-de Sitter spacetime. These are obtained by factoring some
maximally symmetric version by a discrete group of symmetries. The
reference for this material is \cite{kn:gibbons}; see also
\cite{kn:marolf}\cite{kn:khol}\cite{kn:gao2}\cite{kn:lust}\cite{kn:son}
for early work on AdS quotients, and
\cite{kn:emil1}\cite{kn:emil2}\cite{kn:tatar} for more recent
applications of related ideas.

The construction of five-dimensional anti-de Sitter space begins
with the locus
\begin{equation}\label{eq:three}
- A^2 - B^2 + w^2 + x^2 + y^2 + z^2 = -L^2,
\end{equation}
defined in a flat six-dimensional space of signature (2,4). It is
clear that A\2 + B\2 $\geq$ L\2 always, and so circles in this
direction cannot be contracted to a point; on the other hand,
there is no such restriction on the other directions, and we
conclude that the topology of this manifold is S$^1 \times
\bbr^4$. To see what this implies, we choose coordinates defined
such that the time direction is perpendicular to spatial slices.
Such coordinates are given by
\begin{eqnarray} \label{eq:A}
A & = & L\;sin(T/L)                       \nonumber \\
B & = & L\;cos(T/L)cosh(R/L)                       \nonumber \\
w & = & L\;cos(T/L)\; sinh(R/L)\;cos(\chi)                      \nonumber \\
z & = & L\;cos(T/L)\; sinh(R/L)\;sin(\chi)\;cos(\theta)           \nonumber \\
y & = & L\;cos(T/L)\; sinh(R/L)\;sin(\chi)\;sin(\theta)\;cos(\phi)  \nonumber \\
x & = & L\;cos(T/L)\;
sinh(R/L)\;sin(\chi)\;sin(\theta)\;sin(\phi).
\end{eqnarray}
The induced metric is then
\begin{eqnarray}\label{eq:B}
g(AdS_5) & = &  - dT \otimes dT + cos^2(T/L)[dR \otimes dR + L^2 sinh^2(R/L)[d\chi \otimes d\chi \nonumber \\
 & & + sin^2(\chi)\{d\theta \otimes d\theta + sin^2(\theta)d\phi \otimes d\phi\}]].
\end{eqnarray}
With this induced metric, this is a Lorentzian space of negative
curvature $-$1/L$^2$. We see that the spatial sections are copies
of the four-dimensional hyperbolic space H$^4$, with topology
$\bbr^4$. Thus the circle in S$^1 \times \bbr^4$ is timelike,
parametrised by T. The metric resembles a FRW metric with spatial
sections of negative curvature. The apparent singularities (at
intervals of $\pi$L) are coordinate singularities: they occur
because all of the timelike geodesics perpendicular to the spatial
surface at T = constant intersect periodically. Beginning at T =
$-\pi$L, a given collection of timelike geodesics contracts
towards each other, intersecting at T = $-\pi$L/2 and T =
$+\pi$L/2, and the whole cycle repeats after T = $\pi$L is
reached: the period is 2$\pi$L.

It is evident from the formula for the coordinate A that these
coordinates do not cover the entire manifold. Nevertheless, they
do faithfully represent the behaviour of timelike geodesics in
AdS$_5$, and, as Gibbons \cite{kn:gibbons} emphasises, the
periodic intersections of those geodesics is a geometric fact
which cannot be abolished merely by claiming to pass to the
universal cover --- a step which is often said to be necessary to
rid the spacetime of its closed timelike worldlines. In fact, it
is difficult to see how an inertial observer in pure anti-de
Sitter spacetime could determine whether or not the temporal
circles had been ``unwrapped". Time, for him, \emph{is} periodic
as measured by the structures available for his inspection
\cite{kn:gibbons}.

Gibbons advocates a pragmatic attitude: if we are considering some
physical system in anti-de Sitter spacetime which is such that the
natural periodicity of this spacetime is not observable --- for
example, if the (locally measured) period is vast even by
cosmological standards --- then we need \emph{not} pass to the
universal cover. The choice should be determined by the physical
circumstances. In the context of cosmology, spacetimes with
negative cosmological constants typically display spacelike
singularities --- see \cite{kn:MMM} for a discussion of this. In
cosmology, therefore, one can question whether ``unwrapping AdS"
(to its universal cover) is really appropriate. We can think of
the periodicity as a device to avoid pathologies which may arise
in the long run if we introduce matter into anti-de Sitter
spacetime and break time translation invariance.

In our applications, where we regard AdS\5 as a tool to study the
physics of an embedded dS\4 slice, it is possible to argue that the
existence of a cyclic time coordinate on AdS$_5$ is of no physical
concern provided that it can be reconciled with a non-cyclic time
coordinate on the de Sitter slice. In fact, as we
shall see, the de Sitter slice plays out its infinite history
within much less than one ``Great Year" of the ambient anti-de
Sitter spacetime, so here the cyclic version of AdS$_5$ is physically
appropriate in Gibbons' sense. For if we ``unwrap" the ambient
space, most of it will lie ``before or after the infinite past or
future" for de Sitter --- and this seems physically irrelevant.
Furthermore, the proper time of the \emph{static} observers in
anti-de Sitter spacetime is periodic, but not with period 2$\pi$L
in general. In global coordinates (see equation \ref{eq:nine}
below) the proper time for such an observer (at a constant value
of r) is periodic with period 2$\pi$Lcosh(r/L). This implies that,
for static observers in the region where the de Sitter slice
resides, the period of each cycle is enormous. Thus neither de
Sitter observers, nor static bulk observers nearby, would
necessarily be aware of the supposed periodicity of anti-de Sitter
spacetime.

Our attitude will be a conservative one: we take it that the
periodic identification in the timelike direction of the locus
given by equation \ref{eq:three} is just a mathematical device
which helps us to focus on a finite interval of anti-de Sitter
(global) time --- which is all that we shall need. However, it
should be mentioned that some authors are willing to interpret the
periodicity literally, for physical reasons. For example, this has
been discussed by Allen and Jacobson \cite{kn:allen} and recently
by Li \cite{kn:li}. The latter is concerned with a simple de
Sitter brane model devised to explain the smallness of the
observed cosmological constant, and the scheme works best
precisely when the anti-de Sitter bulk is \emph{not} unwrapped to
its universal cover. (The fact that quantum field theory makes
sense on anti-de Sitter spacetime, even when time is cyclic, was
established in \cite{kn:avis}.) Furthermore, the existence of
closed timelike worldlines in AdS-like ``G\"odel" spacetimes has
attracted considerable attention recently in connection with
string dualities \cite{kn:maoz}.

For the sake of definiteness, we shall therefore continue to
define ``AdS$_5$" to be precisely the locus given above, with
topology S$^1 \times \bbr^4$. This ``wrapped" version of anti-de
Sitter is in fact the one being referred to when it is claimed ---
as it usually is --- that the symmetry group is the orthogonal
group O(2,4). (See equation \ref{eq:three}.) For if we unwrapped
this spacetime, then the symmetry group would have to include a
group of time translations with structure $\bbr$, not the O(2)
contained in O(2,4). We stress, however, that even the ``wrapped"
version being considered here is maximally symmetric: one sees
that O(2,4) has the maximal possible dimension (fifteen) for a
five-dimensional semi-Riemannian manifold. (In fact, it is well
known that, in string theory, orientation-reversing isometries of
AdS\5 are not symmetries of the theory, so the precise symmetry is
SO(2,4) rather than O(2,4). In order to avoid confusion, we shall
always take ``symmetry group" to mean the full geometric symmetry
group. It is easy to make the necessary adjustments to preserve the relevant volume form in each case.)

There is still another version of anti-de Sitter spacetime which
is maximally symmetric in this sense, namely the \emph{elliptic
anti-de Sitter spacetime}. The significance of this version can be
understood by means of the AdS/CFT correspondence, as follows.

In the AdS$_5$/CFT$_4$ correspondence, the CFT does not inhabit
Minkowski space, for the conformal group has no natural action
there --- a fact emphasised in \cite{kn:witten2}. Instead it
inhabits conformally compactified Minkowski space CCM$_4$, defined
as the locus (in a flat space of the appropriate signature) given
by coordinates (A,B,w,x,y,z), {\em not all zero}, satisfying
\begin{equation}\label{eq:two}
- A^2 - B^2 + w^2 + x^2 + y^2 + z^2 = 0,
\end{equation}
subject to the identification (A,B,w,x,y,z) = (sA,sB,sw,sx,sy,sz),
where s is any real non-zero number. Following \cite{kn:witten2},
we can think of this space as being obtained from AdS\5 by taking
the locus in equation \ref{eq:three} and imposing scaling
invariance. By using a positive scaling factor, we can send all of
A,B,w,x,y,z to infinity, which is tantamount to letting L\2 tend
to zero. Notice however that, having done this, we can still scale
(A,B,w,x,y,z) to ($-$A,$-$B,$-$w,$-$x,$-$y,$-$z), and so these are
identified on conformally compactified Minkowski space,
\emph{though not on AdS\5 itself}.

The symmetry group of this space is the conformal group PO(2,4).
Here the P refers to the fact that there are {\em two} elements of
the orthogonal group O(2,4) which leave unmoved every point of
CCM$_4$, namely the identity I$_6$ and $-$I$_6$ (since
(A,B,w,x,y,z) = ($-$A,$-$B,$-$w,$-$x,$-$y,$-$z)). Thus the
symmetry group is O(2,4)/\{I$_6$, $-$I$_6$\}, and this by
definition is the \emph{projective} orthogonal group PO(2,4).
Similar considerations show that the topology of CCM$_4$ is that
of $[S^1 \times S^3]/\bbz_2$, which  is very different indeed to
the $\bbr^4$ topology of Minkowski space. A beautiful
interpretation \cite{kn:gibbons} of the difference is given by
regarding Minkowski space as the vector space of 2 $\times$ 2
hermitian matrices --- that is, as the (non-compact) Lie {\em
algebra} of the ``electroweak" group U(2) --- while CCM$_4$ is the
\emph{compact} Lie {\em group} U(2) itself. Note that U(2) is
isomorphic to [U(1) $\times$ SU(2)]/$\bbz_2$, \emph{not} to U(1)
$\times$ SU(2). Thus, CCM$_4$ is the ``matrix exponential of
Minkowski space", and the CFT lives on this matrix exponential.
Note that while it is often claimed that the conformal field
theory of the AdS/CFT correspondence actually inhabits the
universal cover of CCM$_4$, that universal cover does not have
PO(2,4) as its symmetry group. If we are really considering
PO(2,4) (which is usually called ``SO(2,4)" in the literature) to
be ``the conformal group" in the AdS/CFT correspondence, then we
are committing ourselves to a conformal field theory defined on
$[S^1 \times S^3]/\bbz_2$.

The similarity of equation \ref{eq:two} to equation \ref{eq:three}
is apparent, and it is often claimed on this basis that AdS$_5$
and CCM$_4$ have the same symmetry group. That is not the case,
however: the isometry group of AdS$_5$ is, as mentioned earlier,
O(2,4), which is not at all the same as the PO(2,4) symmetry group
of CCM$_4$. In fact, O(2,4) is the double cover of PO(2,4), in
just the same way that SU(2) is the double cover of SO(3). It
therefore follows that, \emph{if indeed} AdS$_5$ \emph{is the
correct version of anti-de Sitter spacetime}, then some states in
AdS$_5$ will be dual to states on CCM$_4$ which are ``spinorial",
in the sense that they will behave non-trivially under apparently
trivial PO(2,4) transformations. (Of course, O(2,4) and PO(2,4)
are identical locally, just as are SU(2) and SO(3). See
\cite{kn:braga} for a strictly local but explicit derivation of
this local isomorphism between the bulk isometry group and the
boundary conformal group.)

We can be more precise about this if we factor AdS$_5$ by the
cyclic group of order two defined by the \emph{antipodal map,}
\begin{equation}\label{eq:six}
\Omega\; : \;(A,B,w,x,y,z)  \rightarrow (-A,-B,-w,-x,-y,-z).
\end{equation}
Call this group $\bbz^{\Omega}_2$. The fixed point of $\Omega$
does not lie on AdS$_5$, so the quotient
AdS$_5$/$\bbz^{\Omega}_2$, which is called EllAdS$_5$, the {\em
elliptic} anti-de Sitter space, is then non-singular and has
exactly the same symmetry group as CCM$_4$, namely O(2,4)/\{I$_6$,
$-$I$_6$\} = PO(2,4). Thus we see that the CFT in the AdS/CFT
correspondence lives not on the boundary of AdS$_5$ but rather on
the boundary of EllAdS$_5$. Indeed, it is EllAdS$_5$ that has
CCM$_4$ as its conformal boundary, as can be seen by observing
that the topology of EllAdS$_5$ is [$S^1 \times \bbr^4$]/$\bbz_2$,
while as we saw the topology of CCM$_4$ is $[S^1 \times
S^3]/\bbz_2$. \emph{It does not, however, follow from this that}
EllAdS\5 \emph{is the ``correct" version of anti-de Sitter
spacetime}. It simply means that, to reach the CFT from AdS\5, we
have to proceed to the conformal boundary \emph{and then project
down to} CCM\4. Since the antipodal map is trivial as far as the
``vector" states on CCM$_4$ are concerned, the CFT can only be
aware of it through ``spinorial" states which transform according
to O(2,4), \emph{not} PO(2,4). On the other hand, if in fact
EllAdS$_5$ is the ``correct" version of anti-de Sitter spacetime,
then such ``spinorial" states will not exist.

The physical meaning of the antipodal map will be discussed
further below. For the present let us note an agreeable property
of EllAdS$_5$: it is both time and space orientable. In fact,
EllAdS$_n$ is always time-orientable \cite{kn:gibbons}, but it is
space-orientable only when n is odd, as is the case here. This is
in contrast to the more familiar \cite{kn:parikh}\cite{kn:aguirre}
elliptic de Sitter space, which is never time-orientable. Notice
that PO(2,4) has the maximum possible dimension for an isometry
group of a five-dimensional manifold, namely 15. Hence EllAdS$_5$
has as much right to be considered ``the maximally symmetric
five-dimensional spacetime of constant negative curvature" as
AdS$_5$.

Thus we see that there are two maximally symmetric versions of
``wrapped" anti-de Sitter spacetime: AdS$_5$ and EllAdS$_5$.
Conformally compactified Minkowski space, CCM\4, is the conformal
boundary of EllAdS\5, and it is reached from AdS\5 simply by
proceeding to the boundary and then projecting down to CCM\4. Both
arrangements are of course compatible with the AdS/CFT philosophy.
If we are willing to break the maximal symmetry, then there are
many other versions as well; but these all descend from AdS$_5$
and EllAdS$_5$, and these descendants will be considered in a
later section. First however let us consider some of the possible
ambiguities of \emph{de Sitter} spacetime.

The simply connected version of four-dimensional de Sitter
spacetime, which we call dS(S$^3$), is given by the locus, in a
space of signature (1,4), defined by
\begin{equation}\label{eq:four}
- A^2 + w^2 + x^2 + y^2 + z^2 = +L^2.
\end{equation}
It is easy to see that the topology is $\bbr \;\times$ S$^3$. The
induced metric is maximally symmetric, with ten-dimensional
isometry group O(1,4), and has constant positive curvature 1/L\2.
The elliptic de Sitter spacetime ElldS$_4$, defined in the obvious
way by antipodal identification, has an isometry group PO(1,4) =
O(1,4)/$\bbz_2$. (This happens to be isomorphic to SO(1,4), which
of course is a subgroup of O(1,4), but the reader should not be
misled by this: the fact that the quotient is isomorphic to a
subgroup is a peculiarity of this case. More typically, PO(2,4) is
certainly \emph{not} isomorphic to SO(2,4).) Of course, ElldS$_4$
is maximally symmetric, so we have the same kind of ambiguity here
as in the anti-de Sitter case. However, when we require a de
Sitter spacetime to fit inside anti-de Sitter spacetime, the two
ambiguities may be said to clash, in the following sense.

The point is that, contrary to what one might expect, EllAdS$_5$
does {\em not} contain elliptic de Sitter space, but rather the
usual dS($S^3$). To see this, take the above equation defining
AdS$_5$ and write it as follows:
\begin{equation} \label{eq:seven}
- A^2 + w^2 + x^2 + y^2 + z^2 = B^2 - L^2.
\end{equation}
Comparing this with the equation defining de Sitter spacetime, we
see at once that for each positive constant B$_0 >$ L there is a
pair of copies of dS($S^3$) (with the same cosmological constant)
embedded in AdS$_5$, one at B = B$_0$, the other at B = $-$B$_0$.
The effect of factoring out $\bbz^{\Omega}_2$ to obtain EllAdS$_5$
is therefore to identify these two copies of dS($S^3$) with each
other: $\bbz^{\Omega}_2$ does not map either copy of dS($S^3$)
into itself. Hence the pair of copies of dS($S^3$) in AdS$_5$
becomes {\em one} copy of dS($S^3$) in EllAdS$_5$, which we take
to be the one at B = B$_0$.

Thus we see that the version of a de Sitter spacetime obtained by
embedding it in \emph{either} AdS$_5$ or EllAdS$_5$ is the simply
connected de Sitter spacetime dS(S$^3$), \emph{not} elliptic de
Sitter. However, this is not to say that the embedding picture
rules out elliptic de Sitter: for we have yet to consider the
consequences of symmetry breaking. We shall return to this below.

Before leaving this discussion, we stress the following point. In
contrast to the Euclidean case, in which there is a copy of $S^4$
passing through every point of $H^5$ except the origin, the
Lorentzian manifolds AdS$_5$ and EllAdS$_5$ are not completely
foliated by copies of dS($S^3$); this only works in the region
$\mid B\mid$ $>$ L. For simplicity, we shall henceforth mainly
focus on this region of AdS$_5$ and EllAdS$_5$.

There is still another ambiguity associated with de Sitter
spacetime; in this case it is related to the conformal
compactification.

It is well known that the de Sitter metric may be written using
conformal time as
\begin{eqnarray}\label{eq:C}
g(dS_4) = {{L^2} \over {sin^2(\eta)}}[- d\eta \otimes d\eta +
d\chi \otimes d\chi + sin^2(\chi)\{d\theta \otimes d\theta +
sin^2(\theta)d\phi \otimes d\phi\}],
\end{eqnarray}
where L\2 is as in equation \ref{eq:four}, and where $\eta$ takes
its values in the open interval (0,$\pi$). We can therefore
compactify by extending conformal de Sitter time to the compact
\emph{closed} interval [0,$\pi$]. This is the usual picture. What
is not generally appreciated, however, is that \emph{this is a
choice}: there is another interpretation, which in some ways is
more natural.

To see this, let us replicate the procedure we used to construct
CCM\4. We use a positive scaling factor to send all of the
coordinates in equation \ref{eq:four} to infinity, which
effectively sends L\2 to zero. The resulting locus
\begin{equation}\label{eq:D}
- A^2 + w^2 + x^2 + y^2 + z^2 = 0,
\end{equation}
with an overall scaling of the coordinates, represents \emph{one}
copy of S\3. To see why this is so, notice that we can use an
overall scaling by a \emph{positive } factor to obtain from
\ref{eq:D}
\begin{equation}\label{eq:DD}
A^2 = 1 =  w^2 + x^2 + y^2 + z^2,
\end{equation}
which represents \emph{two} unit three-spheres, one at A = 1, the
other at A = $-$1. But the remaining freedom of scaling by $-$1
identifies these two.  Thus there is a natural sense in which the
conformal manifold associated with de Sitter spacetime is
\emph{one} copy of S\3. Actually, the conformal representation of
S\3 given by equation \ref{eq:D} is the conformal boundary of
elliptic de Sitter spacetime ElldS$_4$. Exactly as in the anti-de
Sitter case, to reach the conformal space which a CFT might be
expected to inhabit, one proceeds to the boundary and then
projects. The great difference, of course, is that in the de
Sitter case the boundary is disconnected, and the projection
converts two distinct copies of S\3 into one.

We can actually implement this identification \emph{without losing
time-orientability} --- that is, without accepting elliptic de
Sitter as the ``correct" version of de Sitter spacetime --- in the
following novel way. Notice that dropping the conformal factor
from $g$(dS$_4$) in equation \ref{eq:C} has \emph{two} effects:
firstly of course it makes the extension to [0,$\pi$] possible,
but secondly it removes all dependence on $\eta$. Thus
translations along the conformal time axis are symmetries as far
as the conformally deformed metric is concerned, and, in
particular, the map $\eta \rightarrow \eta + \pi $ is now a
symmetry. (Nothing of this sort happens in the anti-de Sitter
case: see equation \ref{eq:ten} below.) We may therefore assume
that the conformal time coordinate $\eta$ values 0 and $\pi$ are
actually identified with each other, so that de Sitter space is
regarded as an open submanifold of a compact space with topology
S$^1$$\times$ S$^3$. We immediately stress that this
identification only affects the unphysical, compactified
spacetime, of which de Sitter spacetime is a proper submanifold;
there are now closed timelike worldlines in the unphysical space,
\emph{but not in de Sitter spacetime itself}. The situation is
clarified by a glance at the Penrose diagram, given in Figure 1.

\begin{figure}[!h]
\centering
\includegraphics[width=0.4\textwidth]{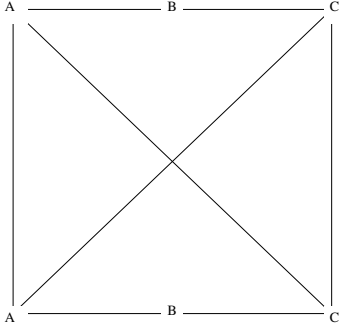}
\caption{Cyclic Compactification of dS($S^3$).}
\end{figure}

Here we have assumed the conventional topology, S$^3$, for the
spatial sections of dS$_4$. The letters at top and bottom indicate
the identification. It is clear that the identification cannot be
detected in a finite amount of proper time by any observer. In
view of this, one might well ask whether there is anything to be
gained from this interpretation of the conformal compactification
of de Sitter spacetime.

The answer is that, \emph{classically}, there is indeed nothing to
be gained. But it is otherwise when we try, following Witten and
Strominger \cite{kn:witten1}\cite{kn:strominger}, to take a
holographic view of de Sitter spacetime. The matching of the bulk
and boundary symmetry groups is an important aspect of the AdS/CFT
correspondence, one which we in fact used above. Contrary to what
is sometimes asserted, however, the isometry group of de Sitter
spacetime does \emph{not} at all match the conformal group of the
boundary, in the usual interpretation of the Penrose diagram. For
it is clear that the boundary in that case is \emph{disconnected},
consisting of two copies of the three-sphere S$^3$. The conformal
group of S$^3$ is SO(1,4) (see equation \ref{eq:D} above, and
recall that O(1,4)/$\bbz_2$ = SO(1,4)); thus the full conformal
group of the S$^3$ $\bigcup$ S$^3$ boundary is the semi-direct
product [SO(1,4) $\times$ SO(1,4)]$\; \triangleleft \; \bbz_2$,
where $\bbz_2$ corresponds to the map which exchanges the two
copies of S\3. This is of course far larger than O(1,4), the
isometry group of dS(S$^3$).

Now Witten and Strominger observe, in discussing the possible
existence of a ``dS/CFT duality", that a scalar field correlator
between a point on the sphere in the infinite past and the
antipodal point (in space and time) on the sphere in the infinite
future is singular. This is to be expected, because the two points
are causally connected by a null geodesic. Strominger argues that,
since the Green functions ``know" about this causal connection,
they only transform simply under \emph{one} copy of the conformal
group SO(1,4); which suggests that the dS/CFT duality should
involve a conformal field theory defined on \emph{one} copy of
S$^3$. An interesting but somewhat drastic way to implement this
suggestion is to modify the geometry of de Sitter space
\emph{itself} so that there is only one boundary component. This
leads to elliptic de Sitter spacetime
\cite{kn:parikh}\cite{kn:aguirre}, and consequently to the loss of
time orientability; this is a very serious drawback, for it
entails all manner of interpretational questions which have not
yet been fully resolved, and it is hard to see how such a
spacetime can be related to conventional FRW models.

A simpler and less drastic alternative, however, is to modify the
\emph{compactification} instead of the spacetime, and this is what
we did above (Figure 1). The interior of the diagram is entirely
unaffected, so time orientability is not lost; nor is causality
affected; nor do we lose the maximal symmetry. (The
Witten-Strominger past-future correlator singularity now becomes a
singularity for an antipodal correlator on S\3, necessitating a
further topological identification which will be discussed below.)

The conformal group of the ``boundary" --- it is no longer a
boundary in the strict topological sense --- is now SO(1,4)
instead of [SO(1,4) $\times$ SO(1,4)]$\; \triangleleft \; \bbz_2$,
while the isometry group of the bulk remains as O(1,4). These are
not exactly the same, just as the symmetry groups of AdS$_5$ and
CCM$_4$ are not exactly the same. As in the previous case, this
means that, if there is some kind of dS/CFT ``correspondence" or
``relationship", then there may be spinorial states on S$^3$ to
account for states on de Sitter spacetime which are not trivial
under the antipodal map --- recall that O(1,4)/$\bbz_2$ = SO(1,4).
Alternatively, if elliptic de Sitter is the correct version, then
these spinorial states will not exist. (The ``missing" $\bbz_2$ is
the one which exchanges future with past. This is an isometry of
all these versions of de Sitter space, but of course it is not a
symmetry which the ``boundary" can be expected to detect after we
have identified future infinity with past infinity.)

Thus, in the de Sitter case, we encounter yet another ambiguity in
the definition of the spacetime, or, rather, of its conformal
compactification. The idea that the de Sitter boundary should
``really" or ``holographically" have one connected component has
been disputed \cite{kn:bala}, and continues to be debated
\cite{kn:kluson}\cite{kn:mann}, but we shall not enter into this
question here; we shall return to it below. The point is that
there is indeed a question, one which we hope to settle by
embedding de Sitter in anti-de Sitter.

Let us summarize the results of this section. Both anti-de Sitter
and de Sitter spacetimes can be interpreted in many ways. In this
section, we have confined ourselves to versions which are
\emph{maximally symmetric}; even so, there are still several
possibilities. For simplicity, we follow Gibbons \cite{kn:gibbons}
and restrict attention to versions of anti-de Sitter spacetime
with cyclic time, since these seem to be the most relevant
versions for dealing with embedded de Sitter spacetimes. Even
then, there are two maximally symmetric versions: AdS$_5$, given
by equation \ref{eq:three}, and its elliptic form,
AdS$_5$/$\bbz^{\Omega}_2$. On the other hand, we found three
maximally symmetric versions of de Sitter spacetime, namely
dS(S$^3$), its elliptic version, and its cyclic version (which is
really a new interpretation of the conformal compactification
rather than of the spacetime itself). In the maximally symmetric
cases we have been considering here, it is relatively
straightforward to determine the relationships between all of
these spacetimes and to fix their symmetry groups. When we wish to
break these symmetries, however, the situation becomes
sufficiently complex that we need a more detailed understanding of
the ways in which these spacetimes and their symmetries fit
together. The next two sections are devoted to the relevant
techniques.

\addtocounter{section}{1}
\section*{3. Touching Infinities}

In this section we shall explore the relationship between an
anti-de Sitter spacetime and an embedded de Sitter spacetime in
more detail.

We can introduce a useful \emph{global coordinate system} for
AdS\5 and EllAdS$_5$ as follows. Referring to equation
\ref{eq:three}, define coordinates (t,r,$\chi, \theta, \phi$) by
\begin{eqnarray}\label{eq:eight}
A & = & L\;sin(t/L)\;cosh(r/L)                       \nonumber \\
B & = & L\;cos(t/L)\;cosh(r/L)                       \nonumber \\
w & = & L\;sinh(r/L)\;cos(\chi)                      \nonumber \\
z & = & L\;sinh(r/L)\;sin(\chi)\;cos(\theta)           \nonumber \\
y & = & L\;sinh(r/L)\;sin(\chi)\;sin(\theta)\;cos(\phi)  \nonumber \\
x & = & L\;sinh(r/L)\;sin(\chi)\;sin(\theta)\;sin(\phi).
\end{eqnarray}
The metric on either AdS$_5$ or EllAdS$_5$ in these coordinates is
\begin{eqnarray}\label{eq:nine}
g(AdS_5) & = &  - cosh^2(r/L) \; dt \otimes dt + dr \otimes dr + L^2 sinh^2(r/L)[d\chi \otimes d\chi \nonumber \\
 & & + sin^2(\chi)\{d\theta \otimes d\theta + sin^2(\theta)d\phi \otimes d\phi\}],
\end{eqnarray}
and this form of the metric is globally valid. Therefore it is
this form of the metric which must be used to determine the nature
of conformal infinity: evidently the latter lies at ``r =
$\infty$". Notice that the spacelike part of the metric is just
the metric on hyperbolic space, $H^4$ (see equation \ref{eq:one}).
As usual we can therefore assume that r $\geq$ 0. On AdS$_5$, the
time coordinate t runs from $-\pi$L to $+\pi$L, after which it
repeats itself. (Compare the formula for A in equations
\ref{eq:eight} with the formula for A in equations \ref{eq:A}.) No
timelike curve can be closed unless t increases by at least
2$\pi$L along it (though of course the proper time elapsed will
not in general be equal to 2$\pi$L). On EllAdS$_5$, where the
points with coordinates (t,r,$\chi, \theta, \phi$) and ($\pi L -
t$,$\:$r,$\:$$\pi - \chi, \pi - \theta, \pi + \phi$) are
identified, it is possible for t to increase by less than 2$\pi$L
along a closed timelike curve ---  the curve A = L$\;$sin(t/L), B
= L$\;$cos(t/L), w = z = y = x = 0, is a closed timelike curve
along which t (which is proper time in this case) only increases
by $\pi$L for each circuit. Notice however that this special curve
does not lie in the region in which we are interested, B $>$ L,
which is foliated by copies of de Sitter . Thus we arrive at the
useful conclusion that in both AdS\5 and EllAdS$_5$, \emph{t must
increase by at least $\pi$L along a closed timelike curve; in
fact, it must increase by at least 2$\pi$L in the region} $\mid
B\mid\; >$ L.

The form of the metric given in equation \ref{eq:nine} is useful
because it allows us to correlate the angular coordinates in the
anti-de Sitter bulk with those on the conformal boundary. For if
we write
\begin{eqnarray}\label{eq:ten}
g(AdS_5) & = & - cosh^2(r/L) \; [dt \otimes dt + sech^2(r/L)\; dr \otimes dr + L^2 tanh^2(r/L)\{d\chi \otimes d\chi \nonumber\\
 &  & + sin^2(\chi)[d\theta \otimes d\theta + sin^2(\theta)d\phi \otimes d\phi]\}]
\end{eqnarray}
then we see that the metric at ``r = $\infty$" is
\begin{eqnarray}\label{eq:eleven}
g(AdS_5, \infty) = - dt \otimes dt + L^2 [d\chi \otimes d\chi +
sin^2(\chi)\{d\theta \otimes d\theta + sin^2(\theta)d\phi \otimes
d\phi\}],
\end{eqnarray}
which is just the standard representative of the conformal
structure on $S^1 \times S^3$ (for AdS\5) or on CCM$_4$ = $[S^1
\times S^3]/\bbz_2$ (for EllAdS\5). Clearly the angular
coordinates $\chi, \theta, \phi$ are the same in the bulk and on
the boundary.

We saw that the region B $>$ L of  AdS\5 or EllAdS$_5$ can be
foliated by copies of dS($S^3$), and we can make this explicit by
choosing new \emph{local} time and radial coordinates, $\tau$ and
$\rho$, as follows. ({\em We retain the angular coordinates}
$\chi, \theta, \phi$.)
\begin{eqnarray} \label{eq:twelve}
A & = & L\;sinh(\tau /L)\;sinh(\rho /L)                       \nonumber \\
B & = & L\;cosh(\rho /L)                       \nonumber \\
w & = & L\;sinh(\rho /L)\; cosh(\tau /L)\;cos(\chi)                      \nonumber \\
z & = & L\;sinh(\rho /L)\; cosh(\tau /L)\;sin(\chi)\;cos(\theta)           \nonumber \\
y & = & L\;sinh(\rho /L)\; cosh(\tau /L)\;sin(\chi)\;sin(\theta)\;cos(\phi)  \nonumber \\
x & = & L\;sinh(\rho /L)\; cosh(\tau
/L)\;sin(\chi)\;sin(\theta)\;sin(\phi).
\end{eqnarray}
Notice that the equation for B enforces B $>$ L if $\rho > $ 0.
The anti-de Sitter metric now becomes
\begin{eqnarray}\label{eq:thirteen}
g(AdS_5) & = & d\rho \otimes d\rho + sinh^2(\rho /L)\;[ -d\tau \otimes d\tau + L^2 cosh^2(\tau /L)\{d\chi \otimes d\chi \nonumber\\
 &  & + sin^2(\chi)[d\theta \otimes d\theta + sin^2(\theta)d\phi \otimes d\phi]\}],
\end{eqnarray}
or
\begin{equation} \label{eq:fourteen}
g(AdS_5) =  d\rho \otimes d\rho + sinh^2(\rho/L)g(dS_4),
\end{equation}
where of course
\begin{eqnarray}\label{eq:fifteen}
g(dS_4) = -d\tau \otimes d\tau + L^2 cosh^2(\tau /L)\{d\chi
\otimes d\chi + sin^2(\chi)[d\theta \otimes d\theta +
sin^2(\theta)d\phi \otimes d\phi]\}
\end{eqnarray}
is the global de Sitter metric. Clearly there is a copy of
dS($S^3$) at any fixed value of $\rho >$ 0 in AdS\5 or EllAdS$_5$.
The cosmological constant of such a de Sitter slice at $\rho$
= c, a constant, is $\Lambda_{dS}$ = +3/[L$^2$sinh$^2$(c/L)], which can be made
very small for suitable choices of c and L; notice that the embedded de
Sitter spacetime need not have the same (magnitude) cosmological
constant as the ambient anti-de Sitter spacetime.

The ratio A/B can be computed in both coordinate systems
introduced in this section, and so we obtain at once
\begin{equation}\label{eq:sixteen}
tan(t/L) =  sinh(\tau /L)\;tanh(c/L).
\end{equation}
From (\ref{eq:fourteen}) and (\ref{eq:fifteen}) we see that $\tau$
sinh(c/L) is proper time for de Sitter; as always in de Sitter
spacetime, it extends from $-\infty$ to $+\infty$, and hence so
must $\tau$. But this corresponds to the interval ($-\pi L/2, \;
+\pi L/2$) for t, which means that de Sitter observers  are
entirely unaware of the cyclic nature of time in AdS\5 and
EllAdS$_5$. (Such cycles require t to increase by 2$\pi$L in this
region.) Furthermore, comparing
the expressions for w in equations \ref{eq:eight} and
\ref{eq:twelve}, we see that
\begin{equation}\label{eq:seventeen}
sinh(r/L) =  sinh(c/L)\;cosh(\tau /L),
\end{equation}
which implies that the minimum value of r on the de Sitter slice
is precisely c. An anti-de Sitter bulk observer who stays in the
region r $>$ c will always experience a time dilation factor of at
least cosh(c/L) (see equation \ref{eq:nine}), and so the period of
the proper time experienced by a bulk observer in the vicinity of
the de Sitter slice is at least
2$\pi\sqrt{\textup{L}^2\;+\;(3/\Lambda_{dS})}$, which is of course
a huge number compared to 2$\pi$L. That is, the possibly cyclic
nature of AdS time is far from apparent even to bulk observers in
the neighbourhood of the de Sitter slice.

Following the philosophy advocated by Gibbons \cite{kn:gibbons},
we conclude that the cyclic version of anti-de Sitter space is the
appropriate one here. Indeed, if we were to unwind AdS\5 or
EllAdS$_5$ to its universal cover, then the de Sitter slice would
be repeated endlessly, and all of the other copies would lie
either ``after the infinite future" or ``before the infinite past"
of a given de Sitter slice, which seems physically meaningless.
Actually we can embed \emph{two} copies of de Sitter spacetime
within the 2$\pi$L cycle of the global anti-de Sitter time
coordinate t; one of these lies in the centre of the range of t,
and the other is (apparently) split into two halves, one ``above"
the slice we are considering here, the other ``below". (In fact of
course the periodicity of t means that these two halves are
joined.) If we follow our philosophy to its logical conclusion, we
should try to eliminate this superfluous copy, since it too lies
``beyond infinite time" for a given de Sitter observer. This will
be important later when we discuss taking the quotient of anti-de
Sitter spacetime by an isometry which shrinks t by a factor of
two.

A still more important point now is this. In equation
\ref{eq:one}, the boundary of H$^5$ is at ``r = $\infty$", and
fixing r therefore severs all contact between infinity and the
copy of S$^4$ at r = c, \emph{since S$^4$ is finite (compact)}.
But in equation \ref{eq:fourteen}, it is not clear that fixing
$\rho$ completely severs contact between anti-de Sitter infinity
and the \emph{Lorentzian} de Sitter slice, since the latter is
itself infinitely large in the time direction.
from which it follows that for fixed $\rho =$ c, the consequence
of letting $\tau$ tend to infinity is that r must tend to
infinity. Thus the infinite nature of de Sitter proper time means
that  \emph{the conformal infinity of the de Sitter slice ($\tau
\rightarrow \pm \infty$) actually lies on the conformal infinity
of AdS\5 or EllAdS$_5$ (r $\rightarrow \infty)$.} Thus, the
Euclidean and Lorentzian pictures of this situation are
fundamentally different.

Now of course the AdS/CFT philosophy is that CFT physics on the
boundary gives a complete account of the interior. One might be
tempted to claim, in view of the fact that the dS boundary lies on
the AdS boundary while the dS bulk inhabits the AdS bulk, that
AdS/CFT imposes a similar holographic equivalence between the bulk
and the boundary of a de Sitter spacetime embedded in anti-de
Sitter spacetime. However, it is not at all clear that we obtain
an \emph{exact equivalence} in this way; in fact we would
\emph{not} expect to do so, since the de Sitter slice is not the
part of the anti-de Sitter bulk which is in direct causal contact
with this part of the anti-de Sitter boundary. This may be related
to the difficulties besetting attempts to establish a full
``dS/CFT correspondence". In fact, the de Sitter slice most
closely resembles the relevant AdS\5 null cones when the constant
c is very small compared to L: but in view of the relation
$\Lambda_{dS}$ = +3/[L$^2$sinh$^2$(c/L)], this would mean that the
de Sitter cosmological constant is very large, which of course is
not the physical case. That is, we expect the dS/CFT relationship
to hold exactly only in the limit of a \emph{large}
$\Lambda_{dS}$. These remarks may well be related to the claim, in
\cite{kn:schaar}, that dS/CFT cannot probe the interior of any
given static patch in de Sitter spacetime.

Thus we have our first lesson from embedding dS\4 as a slice in
AdS\5: some kind of dS/CFT correspondence may well be valid, but
unfortunately it is probably only precise in the unphysical limit
of large cosmological constant.

We now turn to another application of de Sitter embedding.

\addtocounter{section}{1}
\section*{4. Breaking Symmetries with Topology}

AdS\5, EllAdS$_5$ and their conformal boundaries have very large
(15-dimensional) groups of symmetries, and of course one is
interested in breaking these symmetries in some cases. This can be
done in the traditional way by means of vacuum expectation values
of scalar fields \cite{kn:klebanov}. The question of breaking the
specifically \emph{conformal} symmetries on the boundary is also
of much interest \cite{kn:gubser}. But there is another approach
to symmetry breaking.

String theory has taught us that one of the most interesting and
subtle forms of symmetry breaking arises when one takes the
quotient of a manifold by a discrete group \cite{kn:kachru}. The
prototype here is ``Wilson loop symmetry breaking", which arises
on Calabi-Yau compactification manifolds which are not simply
connected --- that is, they have been obtained from simply
connected Calabi-Yau manifolds by taking a quotient by a finite
group of holomorphic isometries. It turns out that the existence
of non-contractible loops on the quotient space allows one to
break gauge symmetries. One might call this general phenomenon
``topological symmetry breaking", since it only works on the
non-simply-connected version of the Calabi-Yau manifold. In a
similar (but subtly different) way, taking the quotient by a
discrete group normally breaks some of the \emph{geometric}
symmetries of a space. (This is not an issue for Calabi-Yau spaces
because their symmetry groups are always very small (finite) in
any case.)

It is possible to take the quotient of anti-de Sitter spacetimes
by discrete groups; there are several motivations for doing so
\cite{kn:marolf}\cite{kn:khol}\cite{kn:gao2}\cite{kn:lust}\cite{kn:son}\cite{kn:emil1}\cite{kn:emil2}.
The idea of taking quotients of \emph{de Sitter} spacetime has
attracted much less attention, but there are strong indications
that this will be necessary, as we shall soon argue. Before doing
so, however, let us clarify the precise way in which taking
quotients by discrete groups breaks geometric symmetries. As we
shall see, there is a subtle difference between this kind of
symmetry breaking and the usual kind.

Suppose that one has a manifold M admitting a group G(M) of
diffeomorphisms (such as isometries, conformal symmetries, and so
on). Let $\Gamma$ be a subgroup of G(M) (which need not act
without fixed points on M) and let N($\Gamma$) be the
\emph{normalizer} of $\Gamma$ in G(M). That is,
\begin{equation}\label{eq:eighteen}
N(\Gamma) = \{g \in G(M) \;\mid \; g\gamma g^{-1} \in \Gamma
\;\;\forall \;\;\gamma \in \Gamma\}.
\end{equation}
(In this work, $\Gamma$ will almost always be either $\bbz_2$ or a
product of copies of $\bbz_2$. Clearly, the normalizer of $\bbz_2$
in any larger group will consist of all those elements of the
larger group which \emph{commute} with the generator of $\bbz_2$.
It turns out, though the argument is less straightforward, that
the same is true for a product of copies of $\bbz_2$ (and also for
$\bbz_4$) in the cases we shall consider. Thus, the reader can
interpret ``normalizer" as ``centralizer" in this work.)

Now N($\Gamma$) contains all those elements of G(M) which descend
to well-defined diffeomorphisms of M/$\Gamma$; for if $m\Gamma$ is
any element of the latter, and $g$ is any element of G(M), then
the definition
\begin{equation}\label{eq:nineteen}
(m\Gamma)g = mg\Gamma
\end{equation}
makes sense if and only if $g$ is an element of N($\Gamma$). But
notice that, with this definition, every element of $\Gamma$
itself has no effect on each element of M/$\Gamma$. Thus the
symmetry group of M/$\Gamma$, which we denote by G(M/$\Gamma$), is
\emph{not} N($\Gamma$) (as is sometimes said) but rather the
quotient N($\Gamma$)/$\Gamma$:
\begin{equation}\label{eq:twenty}
G(M/\Gamma) = N(\Gamma)/\Gamma.
\end{equation}
(Of course, $\Gamma$ is a normal subgroup of N($\Gamma$), so this
quotient is always a group.) Clearly, G(M/$\Gamma$) will in
general be substantially ``smaller" than G(M), and so we can say
that factoring by $\Gamma$ has ``broken" G(M) to G(M/$\Gamma$).
Notice that nothing we have said here requires G to be a group of
isometries or $\Gamma$ to act freely. (Notice too that
G(M/$\Gamma$) is \emph{not} in general naturally isomorphic to a
subgroup of G(M), so this kind of symmetry breaking is not quite
the same as the usual kind, as we mentioned above.)

For example, the normalizer of $\bbz^{\Omega}_2$ (equation
\ref{eq:six}) in O(2,4) is the entire group, O(2,4) itself, and so
the isometry group of the elliptic anti-de Sitter space
AdS$_5$/$\bbz^{\Omega}_2$ is precisely O(2,4)/$\bbz_2$ or PO(2,4),
as we saw. Similarly, the isometry group of S\3, namely O(4),
contains the antipodal map in the form of the matrix
diag($-$1,$-$1,$-$1,$-$1), which is normalized by the whole group;
so the isometry group of the real projective space $\bbr$P\3 is
the projective orthogonal group PO(4) = O(4)/$\bbz_2$.

For an example involving conformal rather than isometric
symmetries, consider the space CCM$_4$ discussed in section 2. If
we wish to consider, as in \cite{kn:gubser}, \emph{non-conformal}
versions of AdS/CFT, then of course we should try to break the
specifically conformal symmetries of the boundary (that is, the
symmetries other than the ordinary isometries). Let us show how to
do this. Recall that CCM$_4$ has the structure [S$^1$ $\times$
S$^3$]/$\bbz_2$, where S$^n$ is the n-sphere. This space has the
semi-Riemannian structure given by equation \ref{eq:eleven}; this
is the Einstein static universe metric, with a seven-dimensional
isometry group given by
\begin{equation}\label{eq:twentyone}
Isom(CCM_4) = [O(2) \times  O(4)]/\bbz_2.
\end{equation}
This group is of course a (small) compact subgroup of the full
(conformal) symmetry group, Conf(CCM$_4$) = PO(2,4). Now CCM$_4$
admits an isometry defined by
\begin{equation}\label{eq:twentytwo}
\aleph \; : \;(A,B,w,x,y,z)  \rightarrow (A,B,-w,-x,-y,-z),
\end{equation}
corresponding to the [O(2) $\times$  O(4)]/$\bbz_2$ element
diag($1, 1, -1, -1, -1, -1$). It is easy to see from equation
\ref{eq:two} that $\aleph$ has no fixed point on CCM\4. (Remember
that, by definition, the point (0, 0, 0, 0, 0, 0) does not lie on
CCM$_4$.) Recalling that ($-$A,$-$B,w,x,y,z) =
(A,B,$-$w,$-$x,$-$y,$-$z), we see that if $\aleph$ generates
$\bbz^{\aleph}_2$, then the quotient manifold has the structure
\begin{equation}\label{eq:twentythree}
CCM_4/\bbz^{\aleph}_2 = S^{1/2} \; \times \; \bbr P^3.
\end{equation}
Here S$^{1/2}$ denotes a circle half the circumference of the
original; that is, a circle modulo the action $\phi  \rightarrow
\phi + \pi$. (This is not the usual ``S$^1$/$\bbz_2$", which is
just a closed interval.) The normalizer of $\bbz^{\aleph}_2$ in
the isometry group [O(2) $\times$  O(4)]/$\bbz_2$ is the whole
group, and so, by equation \ref{eq:twenty}, the isometry group of
CCM$_4/\bbz^{\aleph}_2$ is just
\begin{equation}\label{eq:twentyfour}
Isom(CCM_4/\bbz^{\aleph}_2) = \{[O(2) \times
O(4)]/\bbz_2\}/\bbz^{\aleph}_2 = PO(2) \times PO(4) = O(2) \times
PO(4),
\end{equation}
where we have used the fact that PO(2) = O(2). Since PO(4) =
[SO(3) $\times$ SO(3)] $\triangleleft \;\bbz_2$ is the isometry
group of $\bbr$P$^3$, this result is not very surprising in view
of equation \ref{eq:twentythree}. But now let us ask what happens
to the {\em conformal} symmetry group of CCM$_4$ when we factor by
$\bbz^{\aleph}_2$. The normalizer of $\bbz^{\aleph}_2$ in PO(2,4)
is in fact exactly the isometry group of CCM$_4$, [O(2) $\times$
O(4)]/$\bbz_2$; that is precisely why we are interested in
$\bbz^{\aleph}_2$. For it follows, again from \ref{eq:twenty},
that
\begin{equation}\label{eq:twentyfive}
Conf(CCM_4/\bbz^{\aleph}_2) = O(2) \times PO(4) =
Isom(CCM_4/\bbz^{\aleph}_2).
\end{equation}
Thus, while CCM$_4$ has a conformal group PO(2,4) with eight
generators beyond those of the isometry group (equation
\ref{eq:twentyone}), we now see that CCM$_4$/$\bbz^{\aleph}_2$
{\em has no conformal symmetries} other than its isometries: the
specifically conformal symmetries of CCM$_4$ have all been broken.
Thus we might hope that CCM$_4$/$\bbz^{\aleph}_2$ will play a role
in the study of \emph{non-conformal} bulk/boundary duality,
leading to a different approach to that of \cite{kn:gubser}.

With regard to these examples, we stress again that PO(2,4) is not
a \emph{subgroup} of O(2,4), that $\bbz_2 \; \times \; $[SO(3)
$\times$ SO(3)] $\triangleleft \;\bbz_2$ is not a subgroup of
O(1,4), and so on; in each case, the relevant group is related to
the final symmetry group in the same way that SU(2) is related to
SO(3). As in that case, the consequence may be that certain matter
fields may transform ``spinorially" after the quotient is taken.
We saw examples of this earlier, in discussing the elliptic
versions of both anti-de Sitter and de Sitter spacetimes. The
existence or non-existence of such spinorial states could provide
an intrinsic way of distinguishing such spacetimes from their
quotients. In a later section, however, we shall consider a more
decisive way of doing so.

Now that we have a simple technique for deciding how much symmetry
a quotient space possesses, let us turn to the study of the ``less
symmetric" versions of anti-de Sitter and de Sitter spacetimes,
obtained by taking such quotients. We begin with the de
Sitter-like cases, and return later to the anti-de Sitter cases.

\addtocounter{section}{1}
\section*{5. Less-Symmetric Versions of de Sitter}

The four-dimensional, simply connected version of de Sitter
spacetime, dS(S\3), has the maximal number of Killing vectors
(ten), but it is becoming clear that this large group is probably
not entirely physical. Witten \cite{kn:witten1} emphasises that
if, as has been suggested \cite{kn:banks1}, the Hilbert space of
de Sitter quantum gravity is finite-dimensional, then quantum
gravity must break the de Sitter group, O(1,4), to some much
smaller group which (unlike O(1,4)) has finite-dimensional unitary
representations. This is compatible with the fact that de Sitter
spacetime has no spatial infinity at which de Sitter gauge charges
might be evaluated. \emph{The de Sitter group is not the symmetry
group of quantum gravity in an accelerating universe}. It follows
that ordinary de Sitter spacetime is not the right background for
investigating the true nature of the acceleration.

We have suggested elsewhere \cite{kn:mcinnes} that the correct
version of ``de Sitter spacetime" from this point of view is
obtained simply by taking the spatial sections to be copies of
$\bbr$P\3 --- the antipodally identified version of S\3 ---
instead of S\3. Classically, there is no basis whatever for
preferring dS(S\3) to dS($\bbr$P\3). For our purposes here,
however, there is an important geometric distinction, namely the
fact that the isometry groups are very different. In fact, the
isometry group of dS($\bbr$P\3) is, as we shall show in detail
below, the six-dimensional \emph{compact} group $\bbz_2 \;\times$
PO(4), where as usual PO(n) denotes the projective orthogonal
group. (It may be more useful to express this as $\bbz_2 \;\times$
[(SO(3) $\times$ SO(3)) $\triangleleft \; \bbz_2$], as was done in
\cite{kn:mcinnes}.) Thus, merely by changing the interpretation of
de Sitter spacetime in a way which is classically harmless, we
reduce the relevant symmetry group from one which has no
finite-dimensional unitary representations to one which does.
Physically, the effect of replacing S\3 with $\bbr$P\3 is simply
to reduce the multiplicity of mutually boosted families of de
Sitter observers to one family, since boosts do not commute with
the antipodal map on the spatial sections. This is precisely what
normally happens in cosmology: in generic FRW cosmologies there
is, by construction, a family of observers who are distinguished
by being the observers to whom the Universe appears to be
isotropic. Corresponding to this, the symmetry group of a generic
FRW cosmology is six-dimensional, not ten-dimensional: there are
no boost (or time translation) symmetries in cosmology. Thus, one
could regard the non-trivial topology of the spatial sections of
dS($\bbr$P\3) as a simple way of connecting de Sitter spacetime
with more realistic cosmologies.

The claim is that quantum gravity reduces the size of the symmetry
group (from ten Killing vectors to six), and that this is
implemented formally by the non-trivial topology of the spatial
sections of dS($\bbr$P\3). One can think of this as a way of
mediating between ``observer complementarity" (see for example
\cite{kn:banks3}) and ordinary FRW cosmologies. In the former, the
de Sitter group is reduced to the small group of rotational and
time translation symmetries seen by one single ``static" observer,
and it is said that this is the way in which O(1,4) is replaced by
a subgroup which can describe a finite number of physical states.
However, there is a puzzle here: what has become of the spatial
translation symmetries of de Sitter spacetime, which must still
exist? The answer is that there must be another, complementary
description in which the spatial translations are manifest but the
time translation symmetry (seen by one observer) is not. The
$\bbr$P\3 de Sitter spacetime allows us to reduce O(1,4) not to
the ``static" symmetry group but to the compact subgroup
corresponding to a single \emph{family} of distinguished
(isotropic but not static) observers. This is closer to the
conventional procedures of physical cosmology, in the sense that
in cosmology we normally distinguish \emph{families} of observers
whose worldlines fill the entire spacetime, not individual
observers. Some such ``complementarity" seems to be necessary to
give a complete account of the symmetries of quantum de Sitter
spacetime.

Closely related arguments in favour of replacing S\3 with
$\bbr$P\3 come from other studies of de Sitter entropy. For
example, an ingenious attempt \cite{kn:goheer} to derive the
entropy as entanglement entropy founders precisely because of the
high degree of symmetry of dS(S\3). From yet another point of
view, de Sitter entropy is traditionally \cite{kn:gibhawk} derived
not in de Sitter spacetime itself, but rather in Schwarzschild-de
Sitter spacetime, SdS(S\3), which has a Penrose diagram given in
Figure 2.
\begin{figure}[!h]
\centering
\includegraphics[width=0.3\textwidth]{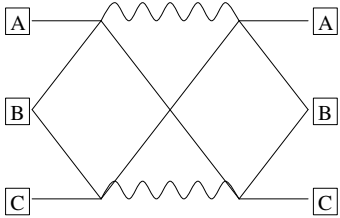}
\caption{Penrose diagram of SdS($S^3$)}
\end{figure}
The left and right sides are topologically identified, as shown.
An attempt to derive the de Sitter entropy formula in terms of
entanglement entropy in this geometry would begin with a pair of
\emph{independent} systems (coupled so as to produce a pure
state), one in each of the diamond-shaped regions in the diagram.
But the ``independence" of those two regions is compromised by the
fact that their future and past infinities are \emph{identical},
due to the topological identifications. One cannot really regard
them as independent if any kind of dS/CFT correspondence is valid,
and we have argued that \emph{some} kind of de Sitter holography
must hold for de Sitter spacetime, even if only very
approximately. The solution to this problem is to note
\cite{kn:mcinnes} that the strange structure of Figure 2 arises
from the assumption that the spatial sections of ``de Sitter
spacetime" have the topology of S\3. If we replace S\3 with
$\bbr$P\3, then SdS(S\3) is replaced by SdS($\bbr$P\3), and it is
shown in \cite{kn:mcinnes} how this splits the conformal infinity
and restores the independence of the two systems.

In view of all this, we shall assume henceforth that the version
of de Sitter spacetime which is most relevant to current
theoretical concerns is one of those which are \emph{not}
maximally symmetric. Let us consider the consequences of replacing
S\3 by $\bbr$P\3 in each of the maximally symmetric versions of de
Sitter spacetime studied earlier.

First, take dS(S\3), given by equation \ref{eq:four}, and consider
the map
\begin{equation}\label{eq:E}
\aleph\; : \;(A,w,x,y,z)  \rightarrow (A,-w,-x,-y,-z).
\end{equation}
(We shall systematically abuse notation and denote by $\aleph$ any
of the maps which reverse the signs of w,x,y, and z but not A or
B.) The fixed points of this map do not lie on the locus, and so
the quotient of dS(S\3) by the $\bbz_2$ generated by $\aleph$ is
non-singular. Clearly we are just performing an antipodal
identification of the spatial sections, leaving time untouched: by
definition, this is the completely non-singular spacetime
dS($\bbr$P\3), whose virtues we have just been describing. The
Penrose diagram of dS($\bbr$P\3) has the form shown in Figure 3.
The stars represent copies of $\bbr$P\2. The detailed
interpretation is given in \cite{kn:mcinnes}.
\begin{figure}[!h]
\centering
\includegraphics[width=0.4\textwidth]{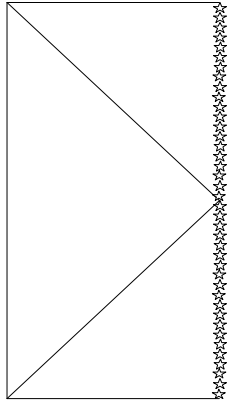}
\caption{Penrose diagram of dS($\bbr{P}^3$).}
\end{figure}

The isometry group of dS(S\3) is O(1,4), and the element of O(1,4)
corresponding to $\aleph$ is diag($1, -1, -1, -1, -1$). This
generates a $\bbz_2$, denoted $\bbz_2^{\aleph}$, which is
normalized by the subgroup $\bbz_2 \; \times \; $O(4), where this
$\bbz_2$ is generated by diag($-1, 1, 1, 1, 1$). Factoring by
$\bbz_2^{\aleph}$, as required by equation \ref{eq:twenty}, we
find that the isometry group of dS($\bbr$P$^3$) is just $\bbz_2 \;
\times \; $PO(4), which may be written as $\bbz_2 \; \times \;
$[SO(3) $\times$ SO(3)] $\triangleleft \;\bbz_2$, where the final
product is semi-direct. ($\bbz_2$ acts by switching the two SO(3)
factors.) Thus topological symmetry breaking can reduce the size
of a symmetry group quite substantially --- in this instance, from
the 10-dimensional non-compact group O(1,4) to the 6-dimensional
compact group $\bbz_2 \; \times \; $[SO(3) $\times$ SO(3)]
$\triangleleft \;\bbz_2$.

Next, recall that we argued (see Figure 1) that it is actually
quite natural to identify the future conformal infinity of dS(S\3)
with its past conformal infinity --- bear in mind that this just
affects the compactification, not the spacetime. However, if we do
that, then the Witten-Strominger past-future correlator
singularity becomes a singularity for an antipodal correlator on
S\3. This is telling us that we ought to perform a further, purely
spatial identification: in other words, the cyclic interpretation
of the Penrose diagram only makes sense physically if, once again,
we systematically replace S\3 with $\bbr$P\3. The effect is that
the compactification of dS($\bbr$P\3) has one boundary at infinity
instead of two, as shown in Figure 4.

\begin{figure}[!h]
\centering
\includegraphics[width=0.4\textwidth]{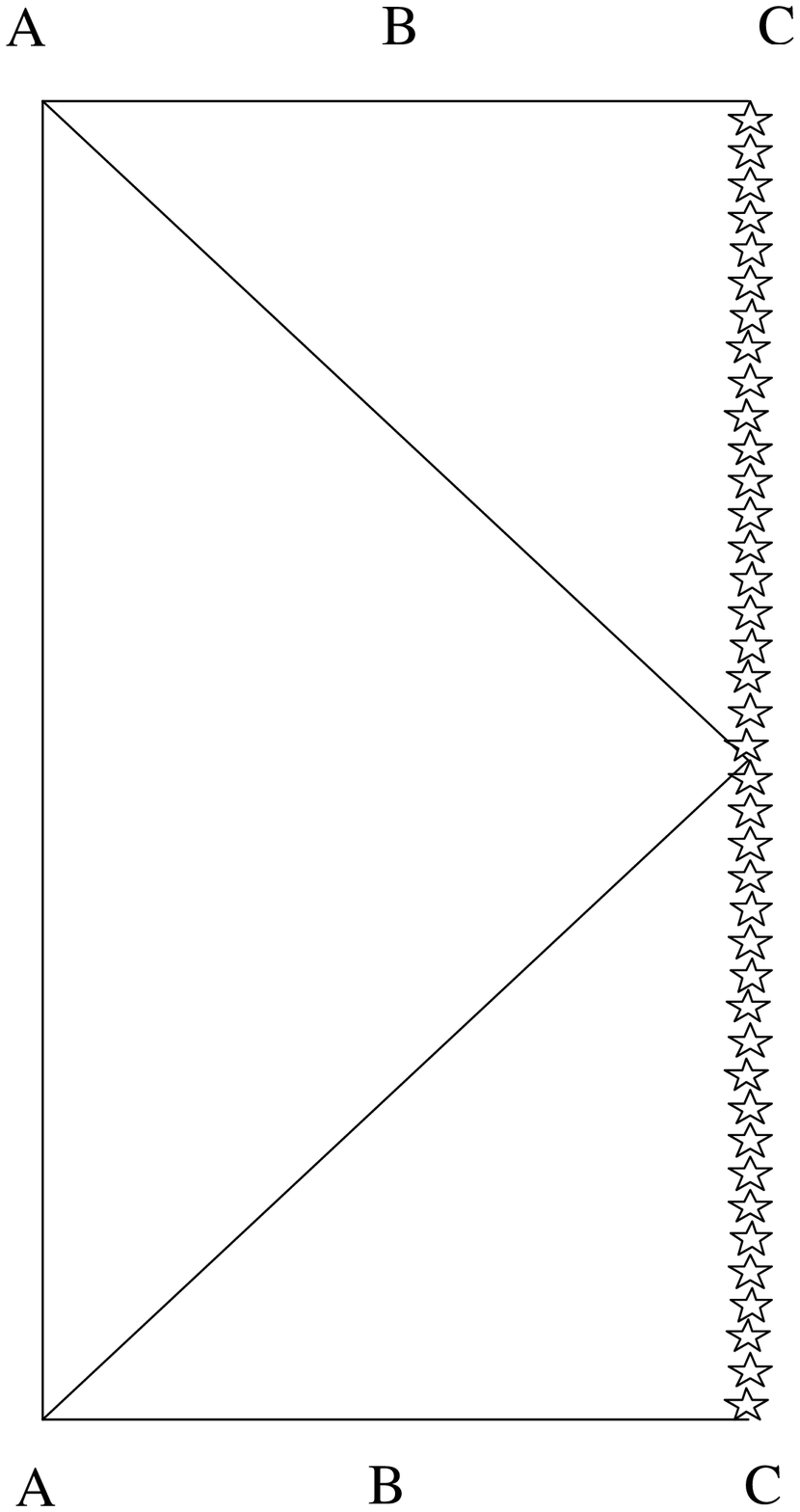}
\caption{Cyclic Compactification of dS($\bbr$P\3)}
\end{figure}
As we have repeatedly stressed, this is a modification of the
conformal compactification, not of dS($\bbr$P\3), so the isometry
group of the spacetime has not changed --- it is still $\bbz_2 \;
\times $ PO(4). However, \emph{the conformal group of the
boundary} has certainly changed. As we saw when discussing Figure
1, the conformal group of S\3 is SO(1,4). By means of another
application of equation \ref{eq:twenty}, one finds that the
conformal group of $\bbr$P\3 is the same as its isometry group ---
it is the projective orthogonal group PO(4). Since both connected
components of the conformal boundary of dS($\bbr$P\3) have the
structure of $\bbr$P\3, we see that the conformal group of the
two-component boundary of dS($\bbr$P\3) is [PO(4) $\times$
PO(4)]$\; \triangleleft \;\bbz_2$, which is not isomorphic to the
isometry group of dS($\bbr$P\3). However, the conformal group of
the ``boundary" of the cyclic compactification is just PO(4),
which is of course far closer to $\bbz_2 \; \times $ PO(4). (They
are still not exactly isomorphic --- as in the case of the cyclic
compactification of dS(S\3), we cannot expect the ``boundary" to
recognise the past/future symmetry of dS($\bbr$P\3).)

Finally we turn to the $\bbr$P\3 version of elliptic de Sitter
spacetime. Note that the full antipodal map on dS(S\3) generates a
$\bbz_2$ which is the diagonal subgroup of the group
$\bbz_2^{\alpha} \times \bbz_2^{\aleph}$ generated by $\aleph$ and
the map
\begin{equation}\label{eq:F}
\alpha\; : \;(A,w,x,y,z)  \rightarrow (-A,w,x,y,z).
\end{equation}
Since the global de Sitter proper time $\tau$ in equation
\ref{eq:fifteen} is related to the coordinate A in equation
\ref{eq:four} by A = Lsinh($\tau$/L), we see that this map just
corresponds to $\tau \rightarrow -\tau$. Extending $\aleph$ to
ElldS\4 in the obvious way, we see that the quotient of ElldS\4 by
$\bbz_2^{\aleph}$ is the same as the quotient
dS(S\3)/[$\bbz_2^{\alpha} \times \bbz_2^{\aleph}$], which simply
means that we identify according to $\tau \rightarrow -\tau$ in
dS($\bbr$P\3). Clearly $\tau \rightarrow -\tau$ has a fixed point
set $\tau$ = 0, which is the $\bbr$P\3 of minimum size in
dS($\bbr$P\3). This is an orbifold singularity in the quotient. We
can think of it as a ``spacelike brane" \cite{kn:strominger2},
which occurs at a finite proper time prior to any point in the
spacetime, and which cuts off the spacetime towards the past.
(Past-directed curves reaching this brane simply terminate there.)

Thus we see that the $\bbr$P\3 version of elliptic de Sitter space
--- let us call it ElldS($\bbr$P\3) ---  differs very greatly from
ordinary ElldS\4: the latter is non-singular but
non-time-orientable, while ElldS($\bbr$P\3) is (orbifold)
singular, but \emph{it is time orientable}. For whereas in ElldS\4
it is not possible to decide globally whether an inextensible
timelike geodesic is future-directed or past-directed, this
\emph{can} be done on ElldS($\bbr$P\3). The reader can confirm
this by following a timelike curve in Figure 5, the Penrose
diagram for this spacetime, as it crosses the diagonal line shown,
and comparing this with the behaviour of a covering curve in
elliptic de Sitter spacetime. (The stars denote copies of
$\bbr$P\2 as usual.)
\begin{figure}[!h]
\centering
\includegraphics[width=0.4\textwidth]{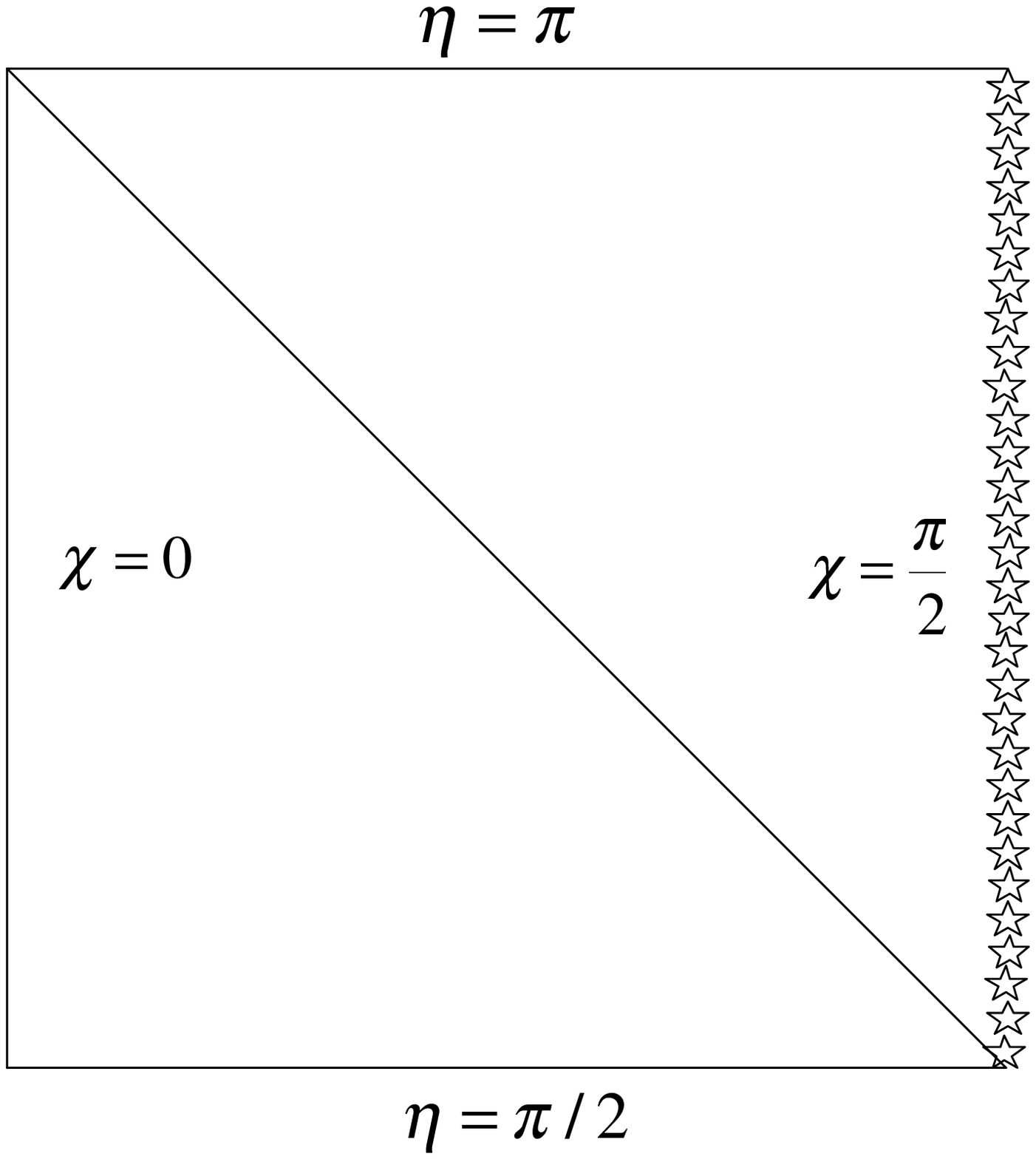}
\caption{ElldS($\bbr$P\3)}
\end{figure}
In elliptic de Sitter, the covering curve changes character from
future-directed to past-directed or vice versa, but  here it does
not do so. In this cosmology, the Universe never contracts: it
begins at the orbifold hypersurface  $\tau$ = 0 (corresponding to
conformal time $\eta$ = $\pi/2$) and expands indefinitely from
there, each spatial section being of course a copy of $\bbr$P\3.
There is only one boundary at infinity, as in elliptic de Sitter
itself; unlike the latter, however, ElldS($\bbr$P\3) has $\bbr$P\3
as its conformal boundary. The normalizer of $\bbz_2^{\alpha}
\times \bbz_2^{\aleph}$ in the dS(S\3) isometry group O(1,4) is
just the same as the normalizer of $\bbz_2^{\aleph}$, namely
$\bbz_2 \; \times \; $O(4), and so using equation \ref{eq:twenty}
as usual we find that the isometry group of ElldS($\bbr$P\3) is
just PO(4), which is of course \emph{precisely} the conformal
group of the $\bbr$P\3 conformal boundary. The reason for this
precise agreement is that ElldS($\bbr$P\3) does not have the
past/future symmetry of the other versions of de Sitter we are
considering here.

Despite its strange character, this cosmology has several
attractive features: the natural absence of curvature
singularities and of any ``bounce" should be noted. Another kind
of expanding cosmology without a big bang is studied in
\cite{kn:ellis}. One could perhaps construct a realistic version
of ElldS($\bbr$P\3) along similar lines, though of course it would
be a challenge to produce the correct ``initial conditions" from
the still poorly-understood physics of spacelike branes. However,
in view of the growing suspicion that the ``de Sitter phase" of
the universe may not be eternal \cite{kn:trivedi}, we may have to
learn to deal with spacetimes of precisely this kind. Notice that
this cosmology is compatible with the very interesting ``final
boundary condition" advocated by Lasenby and Doran
\cite{kn:lasenby}. In short, this kind of cosmology should not be
rejected out of hand. Later we shall argue that it \emph{should}
be rejected, but not for reasons that are obvious at this point.

None of these ``less-symmetric" versions of de Sitter spacetime
can be accommodated in either AdS\5 or EllAdS\5 --- recall that we
saw that the relevant region of the latter was foliated by
dS(S\3). Hence we must modify our versions of anti-de Sitter
spacetime if we wish them to contain a de Sitter slice of one of
these kinds. We shall now show how this is done.

\addtocounter{section}{1}
\section*{6. Less-Symmetric Versions of Anti-de Sitter}

The map $\aleph$ defined by \ref{eq:E} can be extended in the
obvious way from dS(S\3) to AdS\5, via the embedding given by
equation \ref{eq:seven}. (This is formally the same as in equation
\ref{eq:twentytwo}. Alternatively, recall that we have taken care
to use the \emph{same} spherical polar coordinates for anti-de
Sitter space and the de Sitter slice, so $\aleph$ can be extended
in that way, as the map which sends ($\chi, \theta, \phi$) to
($\pi - \chi, \pi - \theta, \pi + \phi$).) The obvious way to
obtain dS($\bbr$P\3) as an AdS slice is to perform this extension
and take the quotient of AdS\5 by the $\bbz_2$ generated by the
extension. Let us see how this works.

While $\aleph$ has no fixed point on dS(S\3), it does have fixed
points on AdS\5 --- it fixes every point on the circle A\2 + B\2 =
L\2, which is the timelike geodesic given in global AdS\5
coordinates by r = 0. In fact this is the worldline of the origin
of these coordinates in the hyperbolic space H$^4$, which, as we
saw (see equation \ref{eq:nine}), gives the geometry of the
spatial sections of AdS\5 in global coordinates. Thus the
extension of $\aleph$ from dS(S\3) to AdS\5 has fixed points,
because the action of $\aleph$ on H$^4$ has a fixed point --- an
important fact to which we shall return later. The quotient
spacetime has the structure
\begin{equation}\label{eq:J}
AdS_5/\bbz_2^{\aleph} = S^1 \;\times [H^4/\bbz_2^{\aleph}].
\end{equation}
It is (orbifold) singular and contains the non-singular spacetime
dS($\bbr$P\3) as a slice (Figure 3). Using equation
\ref{eq:twenty}, one finds that its isometry group is O(2)
$\times$ PO(4). From our discussion of CCM$_4/\bbz^{\aleph}_2$ in
section 5, we know that the relevant conformal group is exactly
the same, O(2) $\times$ PO(4) (see equation \ref{eq:twentyfive}).

There is in fact something rather unsatisfactory about this
procedure, however. We mentioned earlier that each temporal cycle
of AdS\5 contains \emph{two} de Sitter slices --- this can be seen
by comparing the formulae for B in equations \ref{eq:eight} and
\ref{eq:twelve}. (The point is that the coordinates in
\ref{eq:twelve}, which describe one de Sitter slice, require B to
be positive, which, from \ref{eq:eight}, means that t is
restricted to lie between $-\pi$L/2 and $+\pi$L/2. The remaining
range of t allows for another de Sitter slice within the 2$\pi$L
range of t.) The same comments carry over to the quotients we are
considering here: we have in fact embedded dS($\bbr$P\3) in the
(quotient version of) anti-de Sitter twice over. This unappealing
feature can be eliminated as follows.

Define an isometry of AdS\5 by
\begin{equation}\label{eq:P}
\sqrt{\aleph}\; : \;(A,B,w,x,y,z)  \rightarrow (-A,-B,-x,w,-z,y).
\end{equation}
This is of order four, and of course ($\sqrt{\aleph}$)$^2$ =
$\aleph$. Clearly $\sqrt{\aleph}$ does not map the de Sitter slice
in AdS\5 into itself; instead, it maps the slice at $-$B$_0$ to
the one at B$_0$ (see equation \ref{eq:seven}). Having done this,
its square, $\aleph$, converts the S\3 sections to $\bbr$P\3. Thus
we obtain \emph{one} dS($\bbr$P\3) slice in the orbifold quotient
AdS\5/$\bbz_4^{\sqrt{\aleph}}$. (This last is indeed an
\emph{orbifold}, because although $\sqrt{\aleph}$ has no fixed
point, its square does.) This gives us another and more elegant
embedding of dS($\bbr$P\3) as a slice. A somewhat intricate
calculation using the usual techniques shows that the isometry
group of AdS\5/$\bbz_4^{\sqrt{\aleph}}$ is isomorphic to O(2)
$\times$ SO(2) $\times$ SO(3); rather oddly, the bulk has fewer
Killing vectors than the slice. (There is of course no
contradiction in this.)

Turning now to the version of de Sitter pictured in Figure 1 in
``anti-de Sitter" spacetime, we observe that AdS\5 admits an
isometric action by the $\bbz_2$ defined by $\Xi$, given by
\begin{equation}\label{eq:G}
\Xi \; : \;(A,B,w,x,y,z)  \rightarrow (-A,-B,w,x,y,z).
\end{equation}
Unlike $\aleph$, the isometry $\Xi$ has no fixed point on AdS\5,
so the quotient AdS\5/$\bbz_2^{\Xi}$ is actually non-singular. Its
structure is that of S$^{1/2} \times \bbr^4$, where S$^{1/2}$
denotes as usual the circle with half the circumference of the
original one. The shrinking of the circle breaks the O(2,4)
symmetry group of AdS\5 down to O(2) $\times$ O(4); the
corresponding conformal group is the conformal group of
CCM$_4/\bbz^{\Xi}_2$, which is the same as
CCM$_4/\bbz^{\aleph}_2$; hence this group is O(2) $\times$ PO(4).

Like $\sqrt{\aleph}$, $\Xi$ does not map the dS(S\3) embedded in
AdS\5 into itself ; instead it maps the dS(S\3) at $-$B$_0$ to the
one at B$_0$, without changing it.  The effect of factoring out
$\Xi$ is to identify these two slices with each other, leaving
just one de Sitter slice in each cycle of anti-de Sitter global
time.

Now the shrinking of the timelike circle means that the global
anti-de Sitter time coordinate t effectively runs from $-\pi$L/2
to $+\pi$L/2, instead of from $-\pi$L to $+\pi$L. However, we saw
earlier (see equation \ref{eq:sixteen}) that the entire infinite
history of the de Sitter slice extends precisely from t =
$-\pi$L/2 to t = $+\pi$L/2, where the interval can be regarded as
closed \emph{provided} that we include the conformal infinities of
de Sitter spacetime. So in fact the periodicity of global time in
AdS\5/$\bbz_2^{\Xi}$ actually forces the ``time" of the de Sitter
slice to be periodic --- provided of course that this ``time"
extends to the conformal boundary. In other words, it is the
\emph{conformal} de Sitter time $\eta$ (see equation \ref{eq:C})
which is forced to be periodic, \emph{not} the global proper time.
This is of course precisely the situation portrayed in Figure 1.

The picture of the de Sitter slice thus obtained is rather
attractive, since \emph{one} copy of the de Sitter slice sits
neatly in the non-singular ``anti-de Sitter" spacetime
AdS\5/$\bbz_2^{\Xi}$, and the problems associated with the
disconnectedness of the the conformal boundary of de Sitter
spacetime are resolved by the periodicity of anti-de Sitter time.
However, we motivated the picture of the de Sitter
compactification given in Figure 1 by means of the observation
\cite{kn:witten1}\cite{kn:strominger} that a scalar field
correlator between a point on the sphere in the infinite past and
the antipodal point (in space and time) on the sphere in the
infinite future is singular. The idea of Figure 1 was to provide a
geometric formulation of Strominger's argument that the relevant
Green functions only transform simply under \emph{one} copy of the
conformal group SO(1,4); for now there is only one sphere at
infinity. The problem is now to explain the correlator singularity
between antipodal points on this sphere. The obvious step is of
course to replace S\3 by $\bbr$P\3, which is how we obtained
Figure 4. In the present case, the way to obtain the space in
Figure 4 as an AdS slice is simply to extend the definition of
$\aleph$ to AdS\5/$\bbz_2^{\Xi}$ (interpreting equation
\ref{eq:twentytwo} appropriately). The resulting quotient can also
be regarded as a quotient of elliptic anti-de Sitter space by
$\bbz_2^{\aleph}$. The structure is
\begin{equation}\label{eq:H}
EllAdS_5/\bbz_2^{\aleph} = AdS_5/[\bbz_2^{\Xi} \times
\bbz_2^{\aleph}] = S^{1/2} \times [H^4/\bbz_2^{\aleph}],
\end{equation} and it is of course orbifold singular, unlike
AdS\5/$\bbz_2^{\Xi}$. Its isometry group is O(2) $\times$ PO(4),
which is the same as the conformal symmetry group of its boundary,
CCM$_4$/$\bbz^{\aleph}_2$, which was discussed as an example in
section 4.

Finally, we can realise the version of de Sitter pictured in
Figure 5 as an AdS slice by extending the map $\alpha$, defined by
equation \ref{eq:F}, to AdS\5/$\bbz_2^{\aleph}$ by means of the
embedding given by equation \ref{eq:seven}. The effect of course
is just to reverse the anti-de Sitter time coordinates T
(equations \ref{eq:A}) and t (equations \ref{eq:eight}), just as
$\alpha$ reverses the sign of de Sitter time $\tau$. If we take
the quotient AdS\5/[$\bbz_2^{\aleph} \times \bbz_2^{\alpha}$],
then of course we are cutting off T and t at the fixed points,
namely zero and $\pi$L, and placing spacelike branes at the
orbifold singularities there, just as we did in Figure 5. Thinking
in terms of global coordinates, t now extends from zero to $\pi$L,
just as de Sitter global time $\tau$ extends from zero to infinity
in Figure 5. The space pictured in Figure 5 has now been realised
as a slice in AdS\5/[$\bbz_2^{\aleph} \times \bbz_2^{\alpha}$],
which has the structure
\begin{equation}\label{eq:I}
AdS_5/[\bbz_2^{\aleph} \times \bbz_2^{\alpha}] = [0,\pi L] \times
[H^4/\bbz_2^{\aleph}],
\end{equation}
where [0, $\pi$L] is a \emph{closed} interval. The isometry group
of the latter spacetime is computed as follows: the normalizer of
the $\bbz_2^{\aleph} \times \bbz_2^{\alpha}$ group in O(2,4) is
$\bbz_2^{\alpha} \times \bbz_2^{\beta}\; \times $ O(4), where
$\bbz_2^{\beta}$ is generated by the matrix diag(1,$-$1,1,1,1,1).
Therefore the isometry group is $\bbz_2^{\beta} \times$ PO(4).
(There are two copies of the slice for each B$_0 >$ L, one at B =
B$_0$ and one at $-$B$_0$, and the effect of $\bbz_2^{\beta}$ is
just to exchange them. The ``other" de Sitter slice is the
time-reverse of the one at B$_0$, that is, it always contracts; in
the Penrose diagram it sits ``on top of" the given slice. We could
try to eliminate this unwelcome slice in the same way as before,
but we shall not do so here.) The relevant boundary conformal
group is precisely the same. Notice that time is \emph{not}
cyclic, and that there is no continuous time translation (or
``time rotation") symmetry in this version of anti-de Sitter
spacetime.

In this section and the previous one, we have developed a
straightforward, systematic way of analysing some of the
``less-symmetric" versions of [anti-]de Sitter spacetimes and
their relationships. There are of course many others, but the ones
considered here provide a useful, physically interesting sample.
We shall now turn to the problem of using physical arguments to
eliminate some of these spacetimes from contention.

\addtocounter{section}{1}
\section*{7. A Topological Selection Criterion}

We have argued strongly, here and in \cite{kn:mcinnes}, that
dS($\bbr$P\3) (or one of its relatives discussed in the previous
section) is the right version of de Sitter space for
investigations of quantum gravity in an accelerating universe. In
the preceding section, however, we saw that the dS(S\3) isometry
$\aleph$ has fixed points in anti-de Sitter spacetime when it is
extended into the bulk, so that dS($\bbr$P\3) has to be embedded
as a slice in an AdS \emph{orbifold}. The physical importance of
this observation will now be explained.

Adams, Polchinski, and Silverstein \cite{kn:adams} have
conjectured that the condensation of closed string tachyons coming
from the twisted sector of a \emph{non-supersymmetric orbifold}
would tend to resolve the orbifold singularity and restore
supersymmetry. This is implemented by means of a dilaton pulse
which expands outward at the speed of light, ultimately restoring
the geometry to its pre-orbifold state. Strong evidence in favour
of this conjecture has recently been obtained by studying both the
late-time structure \cite{kn:gregory}\cite{kn:headrick} and the
internal consistency of the proposed mechanism
\cite{kn:moore}\cite{kn:zwiebach}. While this work applies
directly to the flat case, it has been argued by Horowitz and
Jacobson \cite{kn:jacob} that a similar phenomenon can be expected
in non-supersymmetric orbifolds of AdS. Indeed, the AdS/CFT
correspondence suggests that matter configurations on slices of
such orbifolds are unstable.

Thus we have an addition to our ``AdS toolbox": if we are obliged
to embed a spacetime as a slice in a non-supersymmetric orbifold
version of AdS, then this is evidence that the object represented
by the slice will not be stable in string theory. As we have seen,
topologically non-trivial versions of de Sitter spacetime do embed
in AdS orbifolds, so the survival of supersymmetry in such
orbifolds gives us a criterion for the acceptability of variant
versions of dS\4.

Now in fact the orbifold singularities of quotients of AdS\5 have
been extensively studied, precisely from the point of view of
supersymmetry breaking. It is convenient to do this by embedding
AdS\5 in a three-dimensional complex flat space $\mathbf{C}^3$, as
follows. Define complex coordinates Z$_1$, Z$_2$, and Z$_3$ in
terms of (A,B,w,x,y,z) by
\begin{eqnarray}\label{eq:K}
Z_1 & = & A + iB                    \nonumber \\
Z_2 & = & w + ix                       \nonumber \\
Z_3 & = & y + iz.                      \nonumber \\
\end{eqnarray}
Then AdS\5 is defined as the locus in $\mathbf{C}^3$ given by
\begin{equation}\label{eq:L}
-Z_1 \overline{Z_1} + Z_2 \overline{Z_2} + Z_3\overline{Z_3} = -
L^2.
\end{equation}
The actions of $\aleph$, $\sqrt{\aleph}$, $\alpha$, and $\Xi$ can
be extended from AdS\5 to $\mathbf{C}^3$ by
\begin{eqnarray}\label{eq:M}
\aleph      & : & (Z_1,Z_2,Z_3) \rightarrow  (Z_1,-Z_2,-Z_3)                   \nonumber \\
\sqrt\aleph & : & (Z_1,Z_2,Z_3) \rightarrow  (-Z_1,iZ_2,iZ_3)                   \nonumber \\
\alpha      & : & (Z_1,Z_2,Z_3) \rightarrow  (-\overline{Z_1},Z_2,Z_3)           \nonumber \\
\Xi         & : & (Z_1,Z_2,Z_3) \rightarrow  (-Z_1,Z_2,Z_3).                      \nonumber \\
\end{eqnarray}
It is immediately clear that the extended action of $\alpha$ on
$\mathbf{C}^3$ is not holomorphic, and we therefore expect that
all supersymmetries are broken in the projection from AdS\5 to
AdS\5/$\bbz_2^{\alpha}$. Therefore, the further quotient
AdS\5/[$\bbz_2^{\aleph} \times \bbz_2^{\alpha}$] (see equation
\ref{eq:I}) has no supersymmetries either. One can in fact prove
this directly simply by recalling our computation of the isometry
group of AdS\5/[$\bbz_2^{\aleph} \times \bbz_2^{\alpha}$], which
is $\bbz_2^{\beta} \times$ PO(4). This group contains \emph{no}
continuous timelike symmetries: that is, the spacetime has no
global timelike Killing vector fields. As is well known from the
de Sitter case (see for example \cite{kn:witten1}), this means
that the spacetime cannot be supersymmetric.

The other three cases are more subtle because $\Xi$, $\aleph$, and
$\sqrt{\aleph}$ do act holomorphically on $\mathbf{C}^3$. Such
actions were analysed in \cite{kn:mukhi}, where the effects of
holomorphic maps on the Killing spinors of AdS\5 were exhibited
explicitly. The results were as follows. (Note that these authors
use the same definition of AdS\5 as we use here, the version
(equations \ref{eq:three} and \ref{eq:L}) with cyclic time.)

Let $\bbz_n$ act on $\mathbf{C}^3$ as follows: if $\gamma$ is a
primitive n\emph{th} root of unity, set
\begin{equation}\label{eq:N}
(Z_1,Z_2,Z_3) \rightarrow  (\gamma^dZ_1,\gamma^aZ_2,\gamma^bZ_3).
\end{equation}
Then Ghosh and Mukhi show that the effect on a general AdS\5
Killing spinor is that of a matrix with eigenvalues
\begin{equation}\label{eq:O}
(a+b-d), -(a+b+d), (a-b+d), -(a-b-d).
\end{equation}
For $\Xi$, we have $\gamma$ = $-$1, d = 1, a = b = 0, and so we
see that none of the eigenvalues is zero; thus neither
AdS\5/$\bbz_2^{\Xi}$ nor any quotient of it, in particular
AdS\5/[$\bbz_2^{\Xi} \times \bbz_2^{\aleph}$], has any
supersymmetry. For $\aleph$, by contrast, we have $\gamma$ = $-$1,
d = 0, a = b = 1, so that precisely two of the eigenvalues vanish,
and we conclude that AdS\5/$\bbz_2^{\aleph}$ has half the
supersymmetry of AdS\5 itself. Finally and most interestingly, for
$\sqrt{\aleph}$ we have $\gamma$ = i, d = 2, a = b = 1, so
AdS\5/$\bbz_4^{\sqrt{\aleph}}$ is quarter-supersymmetric: it too
is a supersymmetric orbifold.

To summarize: of the four versions of anti-de Sitter spacetime we
are considering here, two are supersymmetric (one half, one
quarter) and the other two are not. This means that the version of
de Sitter spacetime pictured in Figure 3 embeds as a slice in the
supersymmetric anti-de Sitter orbifolds AdS\5/$\bbz_2^{\aleph}$
and AdS\5/$\bbz_4^{\sqrt{\aleph}}$, while the version in Figure 4
embeds in the \emph{non-supersymmetric} orbifold
AdS\5/[$\bbz_2^{\Xi} \times \bbz_2^{\aleph}$], and similarly the
version of de Sitter pictured in Figure 5 embeds as a slice in the
non-supersymmetric orbifold AdS\5/[$\bbz_2^{\aleph} \times
\bbz_2^{\alpha}$].

Before drawing any conclusions from this, we should ask whether
the orbifold singularities here are \emph{generic}, in the
following sense. An accelerating universe does not of course have
the exact de Sitter metric: there will be perturbations. The same
is true of the bulk space. Might these perturbations themselves
actually remove the singularities?

We know that the spatial sections of AdS\5 in global coordinates
are just copies of the hyperbolic space H$^4$ (equation
\ref{eq:nine}), which has a very special structure: the sectional
curvatures are all \emph{exactly the same}, independent of both
direction and position. This property would not survive even a
small perturbation. Must a finite group have fixed points on a
perturbed version of H$^4$? Surprisingly, there is a very precise
answer to this difficult question, given by a classical theorem of
Cartan (see \cite{kn:kobayashi}, page 111):
\medskip
\bigskip

THEOREM (Cartan, 1929): Let M be a geodesically complete, simply
connected Riemannian manifold of non-positive sectional curvature,
and let G be a compact group of isometries of M. Then there is a
point in M which is fixed by every element of G.

\bigskip

The key point here is that the theorem does \emph{not} require
that the sectional curvatures should all be the same in all
directions or at all points: it only requires that they should be
non-positive. A small perturbation of H$^4$ will not preserve the
constancy of the curvature, but nor will it change a negative
sectional curvature to one which is positive. Thus, all of the
finite (hence of course compact) groups of isometries we are
considering here, acting on a mildly perturbed version of H$^4$,
will still have a fixed point. The singularity can only be
resolved if the disturbance of the geometry is so large that at
least one sectional curvature reaches a positive value. This has
two consequences: first, the orbifold singularities are indeed
generic, not a result of the highly symmetric geometry of (exact)
AdS\5. Second, if indeed the Adams-Polchinski-Silverstein process
\cite{kn:adams} does resolve the singularities, it can only do so
by means of major disturbances of the geometry, which we can
interpret as stringy instabilities for the relevant versions of de
Sitter spacetime, embedded as slices in the orbifold.

We conclude that the versions of de Sitter spacetime pictured in
Figures 4 and 5 are unstable in string theory, because the
corresponding anti-de Sitter bulk spacetimes are
non-supersymmetric orbifolds. (Before finally abandoning
AdS\5/$\bbz_2^{\Xi}$, however, we note that, while it has no
supersymmetries, nor does it have any orbifold singularities. As
the relevant boundary theory has no conformal symmetries other
than its isometries (both groups are isomorphic to O(2) $\times$
PO(4)), this version of AdS\5 may be of interest for other
purposes, in the study of supersymmetry and conformal symmetry
breaking.)

The simplest non-maximally-symmetric version of de Sitter,
dS($\bbr$P\3), is however still a candidate, for the corresponding
AdS orbifolds are supersymmetric. The examples chosen for
discussion here do not exhaust the list of possibilities, but
precisely the same methods apply in other cases. For example, it
is not hard to show that \emph{elliptic} de Sitter spacetime
embeds in a non-supersymmetric AdS\5 orbifold. We therefore
predict that it is not stable in string theory. In fact, \emph{all
versions of de Sitter spacetime with only one boundary component
are ruled out}. Thus it seems that we must accept that de Sitter
spacetime is \emph{not} ``really" or ``holographically" dual to a
single CFT inhabiting one boundary space. This supports the
version of de Sitter holography put forward in \cite{kn:bala},
with its two independent but entangled CFTs.

The only survivors now are dS($\bbr$P\3) and some of the
generalisations of it obtained by replacing $\bbz_2$ by some
larger finite subgroup of the isometry group of S\3, that is,
O(4). All of these can be obtained as slices in AdS\5 orbifolds;
most cannot be obtained as slices in \emph{supersymmetric} AdS\5
orbifolds, however, so many candidates can be winnowed out; for
example, it can be shown that the recently proposed cosmology with
``dodecahedral" spatial geometry \cite{kn:uzan} would be unstable
in string theory. This is a striking example of the use of string
theory to constrain spacetime topology. Note that it has been
claimed that observations do indeed rule out the ``dodecahedral"
model \cite{kn:cornish}.

In fact it is possible to prove that the only topologically
non-trivial versions of de Sitter spacetime which \emph{do}
survive our criterion consists of the versions with S\3 replaced
by S\3/$\bbz_n$, where $\bbz_n$ acts in such a way that the
quotient is homogeneous (that is, it has a transitive group of
isometries). These are obtained by defining the action of $\bbz_n$
by means of the map \ref{eq:N}, where $\gamma$ is a primitive
n\emph{th} root of unity, and d = 0, a = b = 1. This of course
includes dS($\bbr$P\3) as a special case; all of the corresponding
AdS\5/$\bbz_n$ quotients are half-supersymmetric. These
cosmologies, with n $>$ 2, are distinguished from dS($\bbr$P\3) in
two ways. First, we saw that dS($\bbr$P\3) can be obtained in a
particularly satisfactory way as a slice in
AdS\5/$\bbz_4^{\sqrt{\aleph}}$; recall that this embedding
dispenses with the physically meaningless ``second slice". It is
not hard to see that no such construction is possible for n $>$ 2
(because the relevant isometry of AdS\5 does not exchange the two
slices). Secondly, it is well known (see for example
\cite{kn:mcinnes}) that dS($\bbr$P\3) is the \emph{only}
non-trivial spatial quotient of dS(S\3) with spatial sections
which are \emph{globally} isotropic. Currently \cite{kn:cornish}
there is no evidence for any topologically-induced anisotropies in
our universe. (Notice too that anisotropic quotients considerably
reduce the rotation group seen by the static observers, whose
observations are so crucial for ``observer complementarity"
\cite{kn:banks3}; for the specific cosmologies we are considering
here, of the form dS(S\3)/$\bbz_n$, n $>$ 2, they will only see a
two-parameter group of symmetries, one for time and one for
rotations about an axis.) We conclude very tentatively that, both
theoretically and observationally, dS($\bbr$P\3) is the favoured
version of de Sitter spacetime. The globally anisotropic but
homogeneous versions should however be investigated more
thoroughly.

As far as anti-de Sitter spacetime is concerned, we have found
that the favoured versions are  AdS\5/$\bbz_2^{\aleph}$, and,
perhaps even more so, AdS\5/$\bbz_4^{\sqrt{\aleph}}$. The possible
importance of such quotients of AdS\5 was in fact suggested by
Ghosh and Mukhi \cite{kn:mukhi} on entirely different grounds,
namely an analogy between the AdS\5/$\bbz_2^{\aleph}$ orbifold and
the one obtained by taking the quotient S$^5$/$\bbz_2$, which is
constructed by replacing the obvious S\3 submanifolds by copies of
$\bbr$P\3. This S$^5$/$\bbz_2$ orbifold is of interest because
blowing up its circle of fixed points is a relevant deformation in
the AdS/CFT context. As suggested in \cite{kn:mukhi}, this
indicates that AdS\5/$\bbz_2^{\aleph}$ may have some special role
to play, independently of its role as the bulk corresponding to
dS($\bbr$P\3). Similar remarks apply to
AdS\5/$\bbz_4^{\sqrt{\aleph}}$. The special ``light states"
arising  from flat gauge connections on these versions of AdS\5
\cite{kn:jacob} will undoubtedly be important in understanding
this more completely.

\addtocounter{section}{1}
\section*{8. Conclusion}

Anti-de Sitter spacetime arises so naturally in string theory that
it seems puzzling that the cosmological constant of our world is
positive rather than negative. Perhaps the real role of AdS,
however, is as a tool which can be used to extract answers to
questions which cannot yet be approached using string theory
directly. This is how it has been used \cite{kn:maldacena} to
investigate the stringy status of the black hole information
paradox. As Gibbons observes \cite{kn:gibbons}, one of the most
fundamental questions in quantum gravity is that of how to
translate non-trivial spacetime geometry and topology into the
quantum-mechanical context. We have suggested here that, once
again, embedding in a version of AdS is the way to bring string
theory to bear on this problem.

Quantum gravity in de Sitter spacetime \cite{kn:witten1} suggests
that the relevant version of de Sitter spacetime is one of the
many versions which are not maximally symmetric, and we interpret
this to mean that the physical version is topologically
non-trivial. The question is then: how complex can the topology of
``de Sitter spacetime" become? The answer we have proposed here,
using the ``AdS toolbox", is, ``not very." We saw that
dS($\bbr$P\3), the version advocated by de Sitter himself
\cite{kn:desitter}, explored in \cite{kn:louko}, and discussed in
this context in \cite{kn:mcinnes}, seems to be the natural
candidate for the ``true" form of de Sitter spacetime. However,
more complex quotients of S\3 are still allowed: these are the
versions of de Sitter spacetime which are homogeneous but only
\emph{locally} isotropic. Since $\bbr$P\3 is perfectly isotropic,
even globally
--- unlike any other quotient of S\3 --- recent results
\cite{kn:cornish}, which are consistent with an absence of
topologically-induced cosmic anisotropies, may be said to support
this candidate. But much remains to be done in the effort to
understand whether and why dS($\bbr$P\3) is really preferred to
all other spatially homogeneous versions of de Sitter spacetime.
\addtocounter{section}{1}
\section*{Acknowledgement}
The author is grateful to Wanmei for everything, including the
diagrams in this paper.

\end{document}